\documentclass[final,5p,times,twocolumn]{elsarticle}

\usepackage{amssymb}

\usepackage{lineno}

\usepackage{amsmath,amsfonts}
\usepackage{makecell}
\usepackage{orcidlink}
\usepackage{pifont}
\usepackage{diagbox}
\usepackage{booktabs}

\usepackage{soul}
\sethlcolor{white}

\usepackage{multirow}
\usepackage{tikz}
\usetikzlibrary{arrows.meta, positioning, calc}

\journal{Information Fusion}

\begin{document}

\begin{frontmatter}



\title{On the Security and Privacy of Federated Learning: A Survey with
Attacks, Defenses, Frameworks, Applications, and Future Directions} 


\author{Daniel M.
Jimenez-Gutierrez} 
\author{Yelizaveta
Falkouskaya} 
\author{José L.
Hernandez-Ramos} 
\author{Aris
Anagnostopoulos} 
\author{Ioannis
Chatzigiannakis} 
\author{Andrea Vitaletti} 

\affiliation{organization={Department of Computer, Control and Management Engineering, Sapienza University of Rome},
            addressline={Via Ariosto, 25}, 
            city={Rome},
            postcode={00185}, 
            state={Rome},
            country={Italy}}

\begin{abstract}
Federated Learning (FL) is an emerging distributed machine learning
paradigm enabling multiple clients to train a global model
collaboratively without sharing their raw data. While FL enhances data
privacy by design, it remains vulnerable to various security and privacy
threats. This survey provides a comprehensive overview of more than 200
papers regarding the state-of-the-art attacks and defense mechanisms
developed to address these challenges, categorizing them into
security-enhancing and privacy-preserving techniques. Security-enhancing
methods aim to improve FL robustness against malicious behaviors such as
byzantine attacks, poisoning, and Sybil attacks. At the same time,
privacy-preserving techniques focus on protecting sensitive data through
cryptographic approaches, differential privacy, and secure aggregation.
We critically analyze the strengths and limitations of existing methods,
highlight the trade-offs between privacy, security, and model
performance, and discuss the implications of non-IID data distributions
on the effectiveness of these defenses. Furthermore, we identify open
research challenges and future directions, including the need for
scalable, adaptive, and energy-efficient solutions operating in dynamic
and heterogeneous FL environments. Our survey aims to guide researchers
and practitioners in developing robust and privacy-preserving FL
systems, fostering advancements safeguarding collaborative learning
frameworks' integrity and confidentiality. 
\end{abstract}




\end{frontmatter}



\section{Introduction}
\label{sec:intro}
Machine Learning (ML) has revolutionized numerous
fields~\cite{macnish2025application} by enabling computers to learn from
data and make informed decisions without being explicitly programmed for
every scenario. This capability has become increasingly crucial in
today's data-driven world, where the volume, velocity, and variety of
information far exceed human capacity for manual analysis. ML
applications span a wide range of industries, including
healthcare~\cite{ganatra2025machine}, finance~\cite{biju2024examining},
manufacturing~\cite{jin2024big}, and
entertainment~\cite{guste2024machine}. It offers solutions to
previously intractable problems and opens new frontiers for
innovation. As organizations and researchers seek to leverage the
power of ML, they often face challenges related to data accessibility
and \emph{privacy concerns}.

Federated Learning (FL)~\cite{mcmahan2017communication} has emerged as a powerful paradigm enabling multiple clients (local nodes, parties, participants) to train ML models collaboratively without sharing raw data. While FL enhances data privacy, it also introduces unique \emph{security} and \emph{privacy} challenges that do not exist in traditional centralized learning settings, including vulnerabilities exacerbated by non-IID (non-Independent and Identically Distributed) data, where client datasets exhibit statistical heterogeneity in label, feature, or quantity distributions. Non-IID data amplifies security risks such as poisoning attacks, as adversaries can exploit skewed local updates to manipulate the global model, and privacy risks like membership inference, where attackers infer participation of specific data points by exploiting distributional disparities~\cite{shokri2017membership}. 

The distributed nature of FL makes it vulnerable to various types of attacks, including model poisoning, backdoor attacks, adversarial manipulations, data and gradient leakage, and model update inference, with non-IID conditions further undermining conventional defenses like differential privacy (DP) and robust aggregation. Addressing these challenges is crucial to ensure the robustness, reliability, and trustworthiness of FL systems, especially as they become increasingly adopted in sensitive domains such as healthcare~\cite{kumbhar2025federated, jimenez2023application}, finance~\cite{kong2024asia}, and telecommunications~\cite{salim2025responsible}, among others. 

\subsection{Motivation}
Our survey seeks to present a comprehensive and interconnected overview
of security and privacy in FL. We provide a cohesive perspective on FL's
security and privacy landscape by thoroughly examining various factors
such as attacks, privacy issues, and defense strategies. This integrated
approach enables a deeper comprehension of how FL security and privacy
components are interrelated and influence each other. Through
synthesizing insights from the field, our work aims to offer a complete
understanding of the current state of FL security and privacy, helping
foster a more detailed and nuanced awareness of the challenges and
possibilities in this area.

\begin{table*}[t!]
  \caption{Summary of previous Surveys related to privacy and security in FL (\textcolor{teal}{\ding{52}}: Included, \textcolor{orange}{\ding{117}}: Partially included, \textcolor{purple}{\ding{56}}: Not included)}
  \label{tab:surveys_comparison}
  \resizebox{\textwidth}{!}{%
    \begin{tabular}{cccccccccc}
      \toprule
      \textbf{\makecell{Survey}} & \textbf{\makecell{Publication \\ Year}} & \textbf{\makecell{Security \\ Taxonomy}} & \textbf{\makecell{Privacy \\ Taxonomy}} & \textbf{\makecell{Security \\ Attacks/Defenses}} & \textbf{\makecell{Privacy \\ Attacks/Defenses}} & \textbf{\makecell{Top-tier \\ venues}} & \textbf{\makecell{Frameworks}} & \textbf{\makecell{Fields of \\ Application}} & \textbf{\makecell{Future \\ Directions}} \\
      \midrule
      \textbf{\cite{hu2024overview}} & 2024 & \textcolor{orange}{\ding{117}} & \textcolor{orange}{\ding{117}} & \textcolor{teal}{\ding{52}} & \textcolor{teal}{\ding{52}} & \textcolor{purple}{\ding{56}} & \textcolor{teal}{\ding{52}} & \textcolor{orange}{\ding{117}} & \textcolor{teal}{\ding{52}} \\
      \textbf{\cite{hallaji2024decentralized}} & 2024 & \textcolor{purple}{\ding{56}} & \textcolor{purple}{\ding{56}} & \textcolor{teal}{\ding{52}} & \textcolor{teal}{\ding{52}} & \textcolor{purple}{\ding{56}} & \textcolor{purple}{\ding{56}} & \textcolor{purple}{\ding{56}} & \textcolor{teal}{\ding{52}} \\
      \textbf{\cite{nair2023robust}} & 2023 & \textcolor{orange}{\ding{117}} & \textcolor{orange}{\ding{117}} & \textcolor{orange}{\ding{117}} & \textcolor{orange}{\ding{117}} & \textcolor{purple}{\ding{56}} & \textcolor{purple}{\ding{56}} & \textcolor{purple}{\ding{56}} & \textcolor{teal}{\ding{52}} \\
      \textbf{\cite{neto2023survey}} & 2023 & \textcolor{purple}{\ding{56}} & \textcolor{purple}{\ding{56}} & \textcolor{teal}{\ding{52}} & \textcolor{purple}{\ding{56}} & \textcolor{purple}{\ding{56}} & \textcolor{purple}{\ding{56}} & \textcolor{teal}{\ding{52}} & \textcolor{purple}{\ding{56}} \\
      \textbf{\cite{li2023review}} & 2023 & \textcolor{purple}{\ding{56}} & \textcolor{purple}{\ding{56}} & \textcolor{teal}{\ding{52}} & \textcolor{purple}{\ding{56}} & \textcolor{purple}{\ding{56}} & \textcolor{orange}{\ding{117}} & \textcolor{orange}{\ding{117}} & \textcolor{teal}{\ding{52}} \\
      \textbf{\cite{rodriguez2023survey}} & 2023 & \textcolor{orange}{\ding{117}} & \textcolor{teal}{\ding{52}} & \textcolor{teal}{\ding{52}} & \textcolor{teal}{\ding{52}} & \textcolor{purple}{\ding{56}} & \textcolor{purple}{\ding{56}} & \textcolor{purple}{\ding{56}} & \textcolor{teal}{\ding{52}} \\
      \textbf{\cite{gong2022backdoor}} & 2022 & \textcolor{orange}{\ding{117}} & \textcolor{purple}{\ding{56}} & \textcolor{teal}{\ding{52}} & \textcolor{purple}{\ding{56}} & \textcolor{purple}{\ding{56}} & \textcolor{purple}{\ding{56}} & \textcolor{purple}{\ding{56}} & \textcolor{orange}{\ding{117}} \\
      \textbf{\cite{qammar2022federated}} & 2022 & \textcolor{orange}{\ding{117}} & \textcolor{orange}{\ding{117}} & \textcolor{teal}{\ding{52}} & \textcolor{orange}{\ding{117}} & \textcolor{purple}{\ding{56}} & \textcolor{orange}{\ding{117}} & \textcolor{orange}{\ding{117}} & \textcolor{teal}{\ding{52}} \\
      \textbf{\cite{liu2022threats}} & 2022 & \textcolor{orange}{\ding{117}} & \textcolor{teal}{\ding{52}} & \textcolor{orange}{\ding{117}} & \textcolor{teal}{\ding{52}} & \textcolor{purple}{\ding{56}} & \textcolor{orange}{\ding{117}} & \textcolor{purple}{\ding{56}} & \textcolor{orange}{\ding{117}} \\

      \textbf{\cite{blanco2021achieving}} & 2021 & \textcolor{orange}{\ding{117}} & \textcolor{orange}{\ding{117}} & \textcolor{teal}{\ding{52}} & \textcolor{teal}{\ding{52}} & \textcolor{purple}{\ding{56}} & \textcolor{purple}{\ding{56}} & \textcolor{purple}{\ding{56}} & \textcolor{teal}{\ding{52}} \\
      
      \textbf{\cite{yin2021comprehensive}} & 2021 & \textcolor{purple}{\ding{56}} & \textcolor{teal}{\ding{52}} & \textcolor{purple}{\ding{56}} & \textcolor{teal}{\ding{52}} & \textcolor{purple}{\ding{56}} & \textcolor{purple}{\ding{56}} & \textcolor{purple}{\ding{56}} & \textcolor{teal}{\ding{52}} \\
      \textbf{\cite{mothukuri2021survey}} & 2021 & \textcolor{teal}{\ding{52}} & \textcolor{teal}{\ding{52}} & \textcolor{teal}{\ding{52}} & \textcolor{teal}{\ding{52}} & \textcolor{purple}{\ding{56}} & \textcolor{teal}{\ding{52}} & \textcolor{orange}{\ding{117}} & \textcolor{teal}{\ding{52}} \\
      \textbf{\cite{alazab2021federated}} & 2021 & \textcolor{purple}{\ding{56}} & \textcolor{purple}{\ding{56}} & \textcolor{orange}{\ding{117}} & \textcolor{orange}{\ding{117}} & \textcolor{purple}{\ding{56}} & \textcolor{orange}{\ding{117}} & \textcolor{orange}{\ding{117}} & \textcolor{teal}{\ding{52}} \\
      \textbf{\cite{truong2021privacy}} & 2021 & \textcolor{purple}{\ding{56}} & \textcolor{purple}{\ding{56}} & \textcolor{purple}{\ding{56}} & \textcolor{teal}{\ding{52}} & \textcolor{purple}{\ding{56}} & \textcolor{purple}{\ding{56}} & \textcolor{orange}{\ding{117}} & \textcolor{orange}{\ding{117}} \\
      \textbf{\cite{kholod2020open}} & 2020 & \textcolor{purple}{\ding{56}} & \textcolor{purple}{\ding{56}} & \textcolor{orange}{\ding{117}} & \textcolor{teal}{\ding{52}} & \textcolor{purple}{\ding{56}} & \textcolor{teal}{\ding{52}} & \textcolor{purple}{\ding{56}} & \textcolor{purple}{\ding{56}} \\
      \textbf{Ours} & 2025 & \textcolor{teal}{\ding{52}} & \textcolor{teal}{\ding{52}} & \textcolor{teal}{\ding{52}} & \textcolor{teal}{\ding{52}} & \textcolor{teal}{\ding{52}} & \textcolor{teal}{\ding{52}} & \textcolor{teal}{\ding{52}} & \textcolor{teal}{\ding{52}} \\
      \bottomrule
    \end{tabular}%
  }
\end{table*}

Table~\ref{tab:surveys_comparison} shows a detailed examination of
existing surveys (found following our literature review process
explained in Section~\ref{sec:intro}), revealing significant gaps in
integrating these topics. \textcolor{black}{Despite the growing volume of literature, we observe a fragmented landscape: most prior surveys treat either privacy or security in isolation, often listing threats or defenses without organizing them under a shared conceptual framework. Others omit practical concerns like system frameworks, application domains, or scalability trade-offs.} For example, Hu et al.~\cite{hu2024overview}
and Hallaji et al.~\cite{hallaji2024decentralized} primarily address
security and privacy attacks/defenses but lack coverage of frameworks
and application fields. Similarly, Nair et al.~\cite{nair2023robust},
and Neto et al.~\cite{neto2023survey} offer insights into specific areas
like security defenses or privacy concerns without integrating these
aspects into a broader taxonomy or discussing future directions. Surveys
by Liu et al.~\cite{liu2022threats} and Gong et
al.~\cite{gong2022backdoor} heavily focus on security attacks but do not
comprehensively address privacy or the application of frameworks.

\textcolor{black}{We find that:
\begin{itemize}
    \item Only \textbf{1 out of 16 surveys} attempt to cover both privacy \emph{and} security perspectives.
    \item Fewer than half provide any structured taxonomy of attacks or defenses.
    \item Practical dimensions — such as frameworks and real-world FL applications — are omitted in 12 out of 16 surveys.
    \item None of the existing surveys unify attacks, defenses, and system-level concerns into a single integrated view.
\end{itemize}}

This work aims to build upon and extend the valuable research done in
previous studies by offering a comprehensive and systematized approach
to threats, defenses, and frameworks in FL. We present an extensive
catalog that consolidates and expands upon the diverse sets of threats
and defenses discussed in the existing literature, providing a
multi-faceted categorization of attacks and their corresponding
solutions. Additionally, we examine relevant frameworks, including
privacy and security considerations for FL systems to offer a holistic view of the FL landscape.

\subsection{Contribution}
Our survey addresses this gap by thoroughly reviewing FL's security and
privacy landscape. Table~\ref{tab:surveys_comparison} compares our work
with previous surveys, highlighting our study's unique coverage and
depth. Our survey distinguishes itself by providing a holistic approach
integrating a broad spectrum of critical areas. We cover security and
privacy taxonomies, security and privacy attacks/defenses, and include
discussions on top-tier venues, frameworks, and fields of application.
By offering this comprehensive coverage and systematically describing
attacks from different perspectives, our survey provides a deeper
understanding of the various facets of security and privacy in FL.
Notably, our work is among the few that addresses all these aspects in a
unified framework, thereby offering a complete and cohesive overview for
researchers and practitioners.

Specifically, our contributions are as follows:

\begin{enumerate}
    \item \emph{Comprehensive Taxonomies:} We provide detailed
    taxonomies of security and privacy threats and the corresponding
    defense mechanisms in FL. These taxonomies serve as a structured
    framework for understanding the diverse challenges and solutions in
    the field.
    \item \emph{Inclusion of Frameworks and Applications:} Our survey is
    among the few to cover FL frameworks and real-world fields of
    application for FL. This inclusion offers practical insights into
    how security and privacy measures are implemented and tested in
    real-world scenarios.
    \item \emph{Future Directions and Open Challenges:} We identify
    vital open challenges and outline promising future directions,
    offering valuable guidance for researchers looking to address the
    existing gaps in the literature.
\end{enumerate}

Overall, our survey is distinguished by its broad scope and integrated
approach, making it a valuable resource for researchers and
practitioners seeking a comprehensive understanding of security and
privacy in FL.

\subsection{Relevant Papers Retrieval}

\begin{figure}[!ht]
  \centering
  \includegraphics[width=\linewidth]{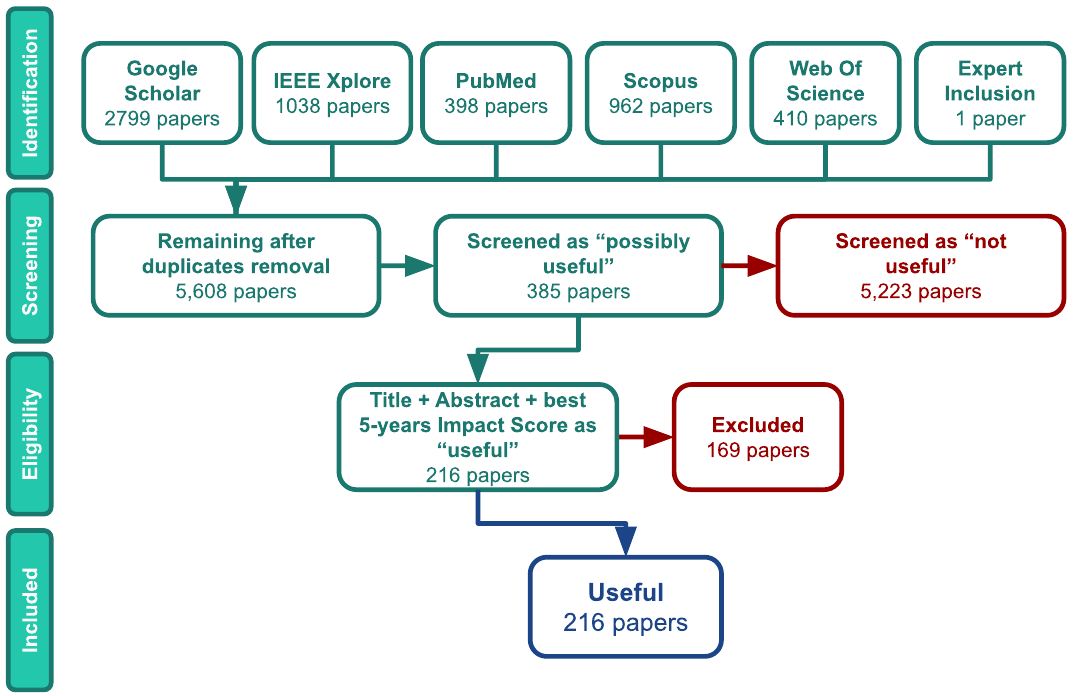}
  \caption{PRISMA flow for gathering relevant references}
  \label{fig:prisma_flow}
\end{figure}

\hl{This work conducted a literature review following the PRISMA methodology}~\cite{siddaway2019systematic} to retrieve and comprehensively analyze FL security and privacy literature. Figure~\ref{fig:prisma_flow} depicts each stage of the methodology to retrieve the most relevant papers. The literature review addressed three key research questions: identifying recent attacks and threats, exploring countermeasures, and evaluating FL frameworks for real-world applications. Using six reputable databases (Google Scholar, IEEE Xplore, PubMed, Scopus, Web of Science), 59 search queries were employed across three themes: attacks, defenses, and
frameworks (see Table~\ref{tab:search_queries}).  After removing duplicates, 2,002 papers were screened based
on keywords, titles, abstracts, and impact scores, ultimately narrowing
the selection to 217 high-quality papers with an impact score above 5.

\begin{table}[h]
    \centering
    \caption{Example search queries by topic}
    \resizebox{\columnwidth}{!}{
    \begin{tabular}{ll}
        \hline
        \textbf{Topic} & \textbf{Example search queries} \\ \hline
        \multirow{3}{*}{Attacks} 
        & ``federated learning attacks'' \\  
        & ``federated learning data poisoning'' \\  
        & ``federated learning backdoor attacks'' \\ \hline
        \multirow{3}{*}{Defenses} 
        & ``federated learning differential privacy'' \\  
        & ``Federated learning secure multiparty computation'' \\  
        & ``federated learning homomorphic encryption'' \\ \hline
        \multirow{3}{*}{Frameworks} 
        & ``federated learning frameworks'' \\  
        & ``Federated learning flower'' \\  
        & ``real-world applications of federated learning'' \\ \hline
    \end{tabular}}
    \label{tab:search_queries}
\end{table}

\subsection{Road Map}

\begin{figure}[htbp]
    \centering
    \begin{tikzpicture}[
        node distance=0.3cm and 0.1cm,
        every node/.style={draw, rounded corners=2pt, minimum width=3.8cm, minimum height=0.7cm, font=\small, align=center},
        arrow/.style={-{Latex[length=2mm]}, thick}
    ]

    \node (intro) {Section 1:\\ Introduction};
    \node (background) [below=of intro] {Section 2:\\ FL Background};

    \node (security) [below=of background, xshift=-2cm] {Section 3:\\ Security Attacks \& Defenses};
    \node (privacy)  [below=of background, xshift=2cm] {Section 4:\\ Privacy Attacks \& Defenses};

    \node (frameworks) [below=of $(security)!0.5!(privacy)$, yshift=-0.8cm] {Section 5:\\ FL Frameworks};
    \node (applications) [below=of frameworks] {Section 6:\\ Applications};

    \node (future) [below=of applications] {Section 7:\\ Future Directions};
    \node (conclusion) [below=of future] {Section 8:\\ Conclusion};

    \draw[arrow] (intro) -- (background);
    \draw[arrow] (background) -- (security);
    \draw[arrow] (background) -- (privacy);
    \draw[arrow] (security) -- (frameworks);
    \draw[arrow] (privacy) -- (frameworks);
    \draw[arrow] (frameworks) -- (applications);
    \draw[arrow] (applications) -- (future);
    \draw[arrow] (future) -- (conclusion);

    \end{tikzpicture}
    \caption{Overview of the paper structure. Each section builds on previous content, progressing from foundational concepts to attack and defense taxonomies, frameworks, applications, and future directions.}
    \label{fig:paper_structure}
\end{figure}
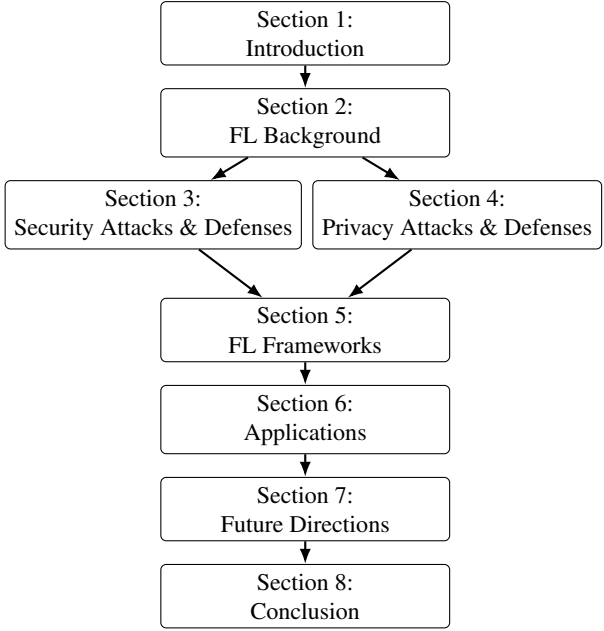

\textcolor{black}{Figure~\ref{fig:paper_structure} outlines the structure of this survey.} In the Section~\ref{sec:fl_backgrnd}, we provide the FL background. 
Section~\ref{sec:security} defines the taxonomy of attacks and defenses
for security in FL. Next, in Section~\ref{sec:privacy}, we provide the
same for privacy in FL. Section~\ref{sec:fl_frameworks} lists the
standardized frameworks for FL. Section~\ref{sec:applications} showcases
the most relevant applications of FL. Then, in
Section~\ref{sec:future_work}, we provide exciting future directions.
Finally, we conclude in Section~\ref{sec:conclusion}.

For the reader's convenience, the acronyms used in this work are listed
in Table~\ref{tab:acronyms}.

\begin{table}[htbp]
    \centering
    \resizebox{\columnwidth}{!}{\begin{tabular}{cc}
    \hline
    \textbf{Acronym} & \textbf{Description} \\ \hline
    ADIs & Adversarial dominating inputs \\ 
    AFR & Anonymous Free-Rider \\ 
    ALIE & A Little Is Enough attack \\ 
    AutoGM & Auto-Weighted GeoMed \\ 
    BFT & Byzantine Fault Tolerance \\ 
    C-GANs & Cross-Client GANs \\ 
    CPA & Cocktail Party Attack \\ 
    DDP & Dynamic Differential privacy \\ 
    DFL & Decentralized federated learning \\ 
    DLG & Deep leakage from gradients \\ 
    DP & Differential Privacy \\ 
    E2EGI & End-to-End Gradient Inversion Attack \\ 
    FC & Fully connected \\ 
    FL & Federated learning \\ 
    FOLTR & Federated online learning to rank \\ 
    FR & Free-Rider \\ 
    GAN & Generative Adversarial Network \\ 
    GDPR & Data Protection Regulation \\ 
    GeoMed & Geometric Median \\ 
    GS & Gradient stalking \\ 
    HIPAA & Health Insurance Portability and Accountability Act \\ 
    HE & Homomorphic Encryption \\ 
    IPM & Inner Product Manipulation \\ 
    IoT & Internet of Things \\ 
    LDP & Local differential privacy \\ 
    MAB & Adversarial Multi-Armed Bandit \\ 
    MarMed & Marginal Median \\ 
    MCS & Mobile crowdsensing \\ 
    MeaMed & Mean Around Median \\ 
    MitM & Man-in-the-Middle \\ 
    ML & Machine learning \\ 
    MPC & Secure Multiparty Computation \\ 
    OT & Oblivious Transfer \\ 
    PASS & Parameter Audit-based Secure and Fair FL Scheme \\ 
    PID & Privacy-aware and incremental defense \\ 
    PMIAs & Poisoning membership inference attacks \\ 
    RoFL & Robustness of secure FL \\ 
    SCA & Sybil-Based Collusion Attacks \\ 
    SFL & Split Federated Learning \\ 
    SFR & Selfish Free-Rider \\ 
    SR & Systematic review \\ 
    SS & Secret sharing \\ 
    TFF & Tensorflow Federated \\ 
    VQA & Visual question-answering \\ 
    ZKP-FL & Zero-knowledge proof-based FL \\ 
    ZKPs & Zero-knowledge proofs \\ \hline
    \end{tabular}}
    \caption{Acronyms employed in this paper}
    \label{tab:acronyms}
\end{table}

\section{FL Background}
\label{sec:fl_backgrnd}

FL~\cite{mcmahan2017communication} is an ML technique for cooperatively
training models on several clients
in a decentralized way, preserving data privacy and ownership for the
client/server owner~\cite{elayan2021deep}. FL is hugely advantageous for
highly decentralized data, especially with the growing prevalence of IoT
devices for continuously capturing data and monitoring users' patterns.

\begin{figure}[!t]
\centering
\includegraphics[width=\columnwidth]{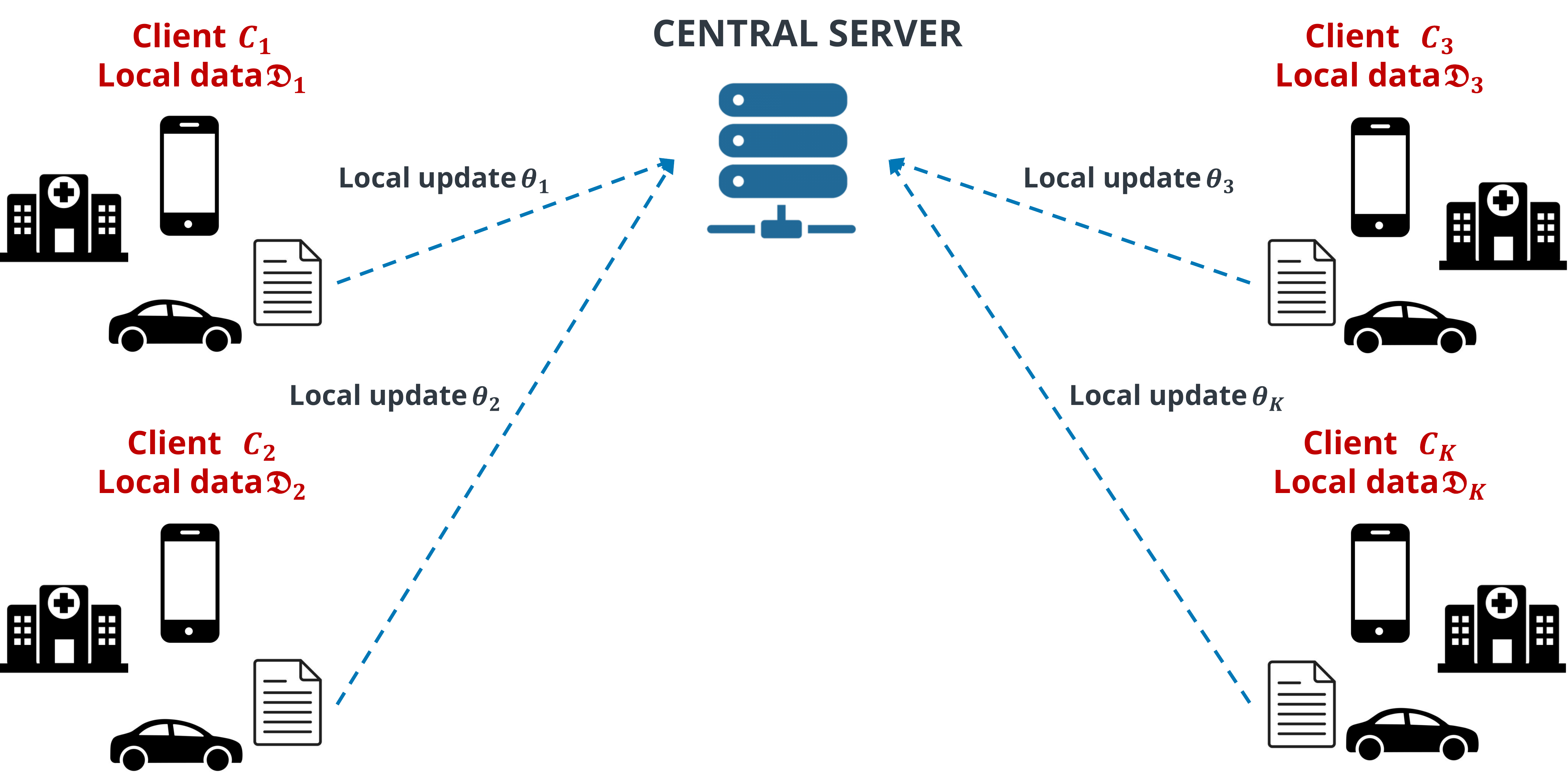}
\caption{FL framework overview}
\label{fig:fl_overview}
\end{figure}

Fig.~\ref{fig:fl_overview} depicts a high-level view of the framework
and how the clients interact with the central server. IoT devices,
institutions (i.e., hospitals, companies, etc.), documents, or vehicles
will collect user data and train a local deep-learning model that
mirrors a previously received global model~\cite{zhang2020federated}.
Following the completion of the local training phase, the models
collaborate to train a global model utilizing their updates rather than
the raw data provided by the users. These model updates indicate changes
in the models' weights during training and do not reflect private or
personal information about the users.

All clients will send updates to a central server,
compiling and using them to aggregate the global model
weights~\cite{sakib2021asynchronous}. Once the global model training
procedure finishes, each client will receive a new copy of the updated
global model. As a result, the models will be trained and updated
regularly without sharing personal information. Thus, the
framework will enable a decentralized architecture in which models get
distributed among clients without requiring a centralized server to
operate the model and serve users. It will also protect users' privacy
by processing and analyzing their data on clients without
disclosing it.

The collaborative model training process in FL involves
\emph{aggregating model updates} from multiple decentralized clients
while preserving data privacy. Aggregation algorithms are pivotal in
this context, serving as the cornerstone for combining these distributed
updates into a global model. These algorithms are essential to ensure
that the federated model achieves the desired convergence and accuracy
while safeguarding the privacy and security of the individual
clients' data. 

\emph{FedAvg}~\cite{mcmahan2017communication} is the most employed
\emph{aggregation algorithm} that operates within a client-server
architecture, where the server orchestrates the training process, and
the clients conduct local training on their data. Each client
independently trains the model using its local data and transmits model
updates to the server. The server aggregates these updates to construct
a global model. FedAvg's advantages include scalability to accommodate a
large user base through decentralized training and improved efficiency
through the ease of computation in a centralized server. However, in FL
settings, one should consider challenges such as client heterogeneity,
communication overhead during update aggregation, and potential
network connectivity limitations.

From a mathematical point of view~\cite{mcmahan2017communication}, FL is
defined by a set of $K$ clients, denoted as $C_1, C_2, ..., C_K$. Each
client $C_i$ has its dataset $\mathcal{D}_i$ containing features
($x$) and labels ($y$) for certain examples (individuals, samples). FL
aims to train a global model $\theta$ in a decentralized manner, where
the model parameters are updated by aggregating the local updates from
each client while keeping the data on the clients. The loss
minimized during the FL  process is $L(\theta) = \sum_{i=1}^{K} (1/K) *
L(\theta_i)$ where $L(\theta)$ is the global loss function to be
minimized and $L(\theta_i)$ is the local loss function for client $C_i$.
This function quantifies the discrepancy between the predictions of the
global model $\theta$ and the ground truth labels for the samples in
client $C_i$'s dataset $\mathcal{D}_i$.

\subsection{Types of FL}
FL is currently in an active development phase and employs diverse
techniques and methodologies to bring its core technology into practical
implementation. When dealing with a nascent technology like FL,
initially categorizing these techniques and approaches is a pivotal
starting point, enabling a more profound comprehension and exploration
beyond the broader conceptual framework. Depending on data partition
and scalability, FL gets divided into different categories
\cite{qammar2022federated,mothukuri2021survey,li2019survey}.

\subsubsection{Data Partition} FL systems are commonly categorized based
on the data distribution across the sample and feature spaces. This
categorization typically divides FL systems into three main types:
horizontal FL, vertical FL, and hybrid FL. Each category represents
distinct approaches to handling data distribution in FL scenarios.

\textbf{Horizontal FL: } In horizontal FL, clients share a common
feature space but have limited overlap in the sample space, making it
suitable for cross-device settings where users collaborate on a shared
task. Local models are trained independently with consistent
architectures, and the global model is updated by averaging local weight
updates.

Mathematically, horizontal FL is represented by contemplating the same
features across clients but with different examples. For example,
suppose clients $C_1$ and $C_2$ have data on different users for a
recommendation system. In that case,
$\mathcal{D}_1 = \{(x_1, y_1), (x_2, y_2), \dots, (x_{n_1}, y_{n_1})\}$
with feature space $X$ and $n_1$ the number of examples of $C_1$, and
$\mathcal{D}_2 = \{(x_1', y_1'), (x_2', y_2'), \dots, (x_{n_2}, y_{n_2})\}$
with feature space $X$ and $n_2$ the number of examples of $C_2$. Here,
$x_i$ and $x_i'$ represent the same features for different
examples~\cite{jimenez2024fedartml,yin2021comprehensive}.

\textbf{Vertical FL: }In vertical FL, datasets from different nodes share the same or similar sample space but differ in the feature space. Entity alignment techniques identify overlapping samples by matching entity descriptions, enabling collaborative training of models like gradient-boosting decision trees. Privacy-preserving methods align entities across clients, facilitating joint gradient training. This approach is often seen in collaborations between different companies.

Mathematically, vertical FL is represented by considering the same set
of examples across clients but with different features. For example, if
clients $C_1$ and $C_2$ have data on patients where $C_1$ has medical
records, and $C_2$ has genetic information, then $\mathcal{D}_1 =
\{(x_1, y_1), (x_2, y_2), \dots, (x_{n_1}, y_{n_1})\}$ with feature
space $X_1$ and $n_1$ the number of examples of $C_1$, and
$\mathcal{D}_2 = \{(x_1', y_1'), (x_2', y_2'), \dots, (x_{n_2}, y_{n_2})\}$
with feature space $X_2$ and $n_1$ the number of
examples of $C_2$. Here, $x_i$ and $x_i'$ represent different feature
sets for the same examples~\cite{yin2021comprehensive}.

\textbf{Hybrid FL: } In numerous other use cases, while conventional FL systems predominantly concentrate on a single type of data partition, the data distribution among the clients often exhibits a hybrid combination of horizontal and vertical divisions. One specific example of this type of FL is \emph{Transfer FL}~\cite{saha2021federated}, which involves horizontal and vertical data partitioning, making it a hybrid approach. The latter allows models to learn from shared features (vertical) and data from different clients (horizontal) to improve performance and generalization.

Let clients $C_1$ and $C_2$ possess datasets $D_1$ and $D_2$ such that:
\begin{equation*}
\begin{aligned}
D_1 &= \bigl\{(x_i^{(1)}, y_i^{(1)})\bigr\}_{i=1}^{n_1},\quad x_i^{(1)} \in \mathcal{X}_1, \\
D_2 &= \bigl\{(x_j^{(2)}, y_j^{(2)})\bigr\}_{j=1}^{n_2},\quad x_j^{(2)} \in \mathcal{X}_2,
\end{aligned}
\end{equation*}
where $\mathcal{X}1$ and $\mathcal{X}2$ are the feature spaces of $C_1$ and $C_2$, respectively. In Hybrid FL, there exist subsets $S{\text{shared}} \subseteq S_1 \cap S_2$ (shared samples) and $X{\text{shared}} \subseteq \mathcal{X}_1 \cap \mathcal{X}_2$ (shared features).

\subsubsection{Scale of Federation} FL fashion can be classified into
two types based on the extent of federation: cross-silo FL and
cross-device FL. The distinctions between these types revolve around the
number of clients and the volume of data stored within
each client.

\textbf{Cross-silo:} The clients are typically organizations or data centers. A limited number of clients are generally involved, each with a substantial volume of data and computational resources. For instance, Amazon aims to offer user-item recommendations by leveraging shopping data from many data centers worldwide~\cite{cheedella2024amazon}.

\textbf{Cross-device:} There is typically a more significant number of clients, each with a comparatively modest amount of data and computational capacity, often consisting of mobile devices. Google Keyboard exemplifies a cross-device FL, where the enhancement of query suggestions in Google Keyboard can benefit from the application of
FL~\cite{wu2024prompt}.

\subsection{Split FL}
Split FL (SFL) is a distributed machine learning approach that utilizes a split model architecture, dividing the model between clients and a central server. This design enhances privacy by avoiding raw data sharing and is suitable for resource-constrained environments due to its distributed computations, which reduce the burden on individual clients~\cite{thapa2022splitfed,thapa2021advancements}. SFL offers high scalability and efficiency in large-scale distributed setups, but comes with limitations, including slower performance compared to traditional FL due to its relay-based training process and increased communication overhead~\cite{singh2019detailed}.

In SFL, for a client $C_i$, the training process proceeds as follows:
\begin{enumerate}
    \item \emph{Forward Pass:} The client computes activations up to the cut layer $a_i = f_{\theta_c}(x_i)$ and sends $a_i$ to the server.
    
    \item \emph{Server Computation:} The server completes the forward pass $\hat{y}_i = f_{\theta_s}(a_i)$.
    \item \emph{Backward Pass:} The server computes the gradient $\nabla_{a_i} \mathcal{L}$ and sends it to the client, which then computes $\nabla_{\theta_c} \mathcal{L} = \nabla_{a_i} \mathcal{L} \cdot \nabla_{\theta_c} f_{\theta_c}(x_i)$.
\end{enumerate}

The overall optimization objective is:
\begin{equation}
    \min_{\theta_c, \theta_s} \frac{1}{K} \sum_{i=1}^K \mathbb{E}_{(x, y) \sim D_i} \left[ \mathcal{L}\left( f_{\theta_s}(f_{\theta_c}(x)), y \right) \right]
\end{equation}

\subsection{Non-IID Data Impact on FL} \label{sec:non-iid-impact}

In FL, non-IID~\cite{g2024noniiddatafederatedlearning} data refers to data that is not uniformly distributed across clients, meaning that different clients may have significantly different data distributions due to factors like user preferences, geographical location, or client usage patterns. Those disparities arise across three dimensions:

\begin{itemize}
    \item \emph{Label Distribution Skew:} Differences in $P(y|x)$ (the conditional probability distribution of labels $y$ given features $x$) between clients. For instance, hospitals specializing in different diseases with imbalanced diagnostic labels.

    \item \emph{Feature Distribution Skew:} Variation in $P(x)$ (the marginal probability distribution of features $x$) across clients. For example, smartphones in different regions capture distinct visual patterns (e.g., urban vs. rural environments).    
    
    \item \emph{Quantity Skew:} Disparities in dataset sizes $n_k$ (where $n\_k = |\mathcal{D}_k|$ denotes the number of samples at client $k$) among clients. For example, IoT devices with varying storage capacities collect unequal data points.
\end{itemize}

Non-IID data poses serious privacy and security challenges as it can make models more vulnerable to inference attacks (e.g., membership and property inference)~\cite{yu2025data} since adversaries can exploit statistical discrepancies to extract sensitive information about client data. Additionally, non-IID data exacerbates the impact of poisoning attacks~\cite{zhao2025fedmp}, where adversarial clients can more effectively manipulate global model updates by injecting biased gradients. On the defense side, traditional DP and robust aggregation methods, such as median or trimmed mean-based aggregation, often lose effectiveness in non-IID settings, as the variability in data distributions can lead to excessive noise or biased
updates. Furthermore, anomaly detection methods~\cite{dong2024fadngs} that rely on outlier detection may struggle to distinguish between natural variations due to non-IID data and actual adversarial behavior.


\section{Security in FL}
\label{sec:security}

\begin{figure*}[!ht]
  \centering
  \includegraphics[width=\linewidth]{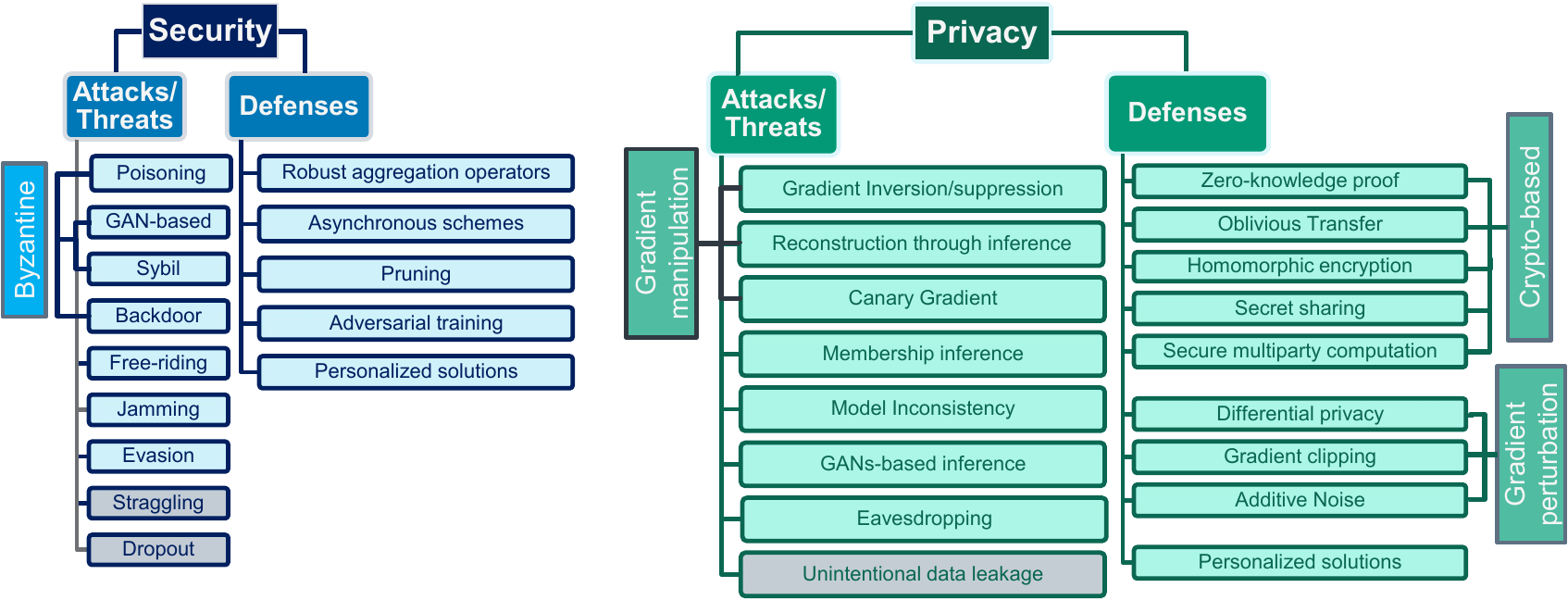}
  \caption{Security and privacy taxonomy for attacks and defenses in FL}
  \label{fig:taxonomy_sec_priv}
\end{figure*}

\hl{%
Based on the papers assessed, we propose a taxonomy of FL security
and privacy attacks and defenses (see
Fig.}~\ref{fig:taxonomy_sec_priv})\hl{, providing a structured
framework for understanding this evolving field. Following such a
taxonomy, we describe the main attacks and defenses for secure FL in
this section. For the attacks, we outline specific mechanisms, degrees
of harm, and specific examples of manifestations in the real world. At
the end of each subsection, we provide some lessons learned after
analyzing the papers regarding attacks and defenses for secure FL.}

\subsection{Security Attacks/Threats}
In FL, security attacks and threats involve adversarial strategies to
compromise models' integrity, availability, confidentiality, and
underlying data. These attacks can be categorized based on several
criteria. \hl{In particular, to enhance clarity and reduce overlaps, we
define five key dimensions for categorizing security attacks and threats
in FL: target specificity, phase affected, intent, nature of the
adversary, and execution style. For target specificity,}
\textit{targeted attacks} aim to disrupt specific system elements, while
\textit{untargeted attacks} seek to cause general disruption or degrade
overall performance. Phase affected clarifies whether the disruption
happens mainly during model training (such as poisoning or Sybil
attacks) or only becomes relevant at inference time (like evasion).
Furthermore, attacks are categorized by intent; \textit{malicious
attacks} aim to cause harm, whereas \textit{exploitative attacks} seek
personal gain without direct harm. Additionally, the nature of the
adversary plays a crucial role: \textit{insider attacks} come from
within the system, while \textit{outsider attacks} originate from
outside~\cite{nair2023robust}. \hl{Finally, execution style clarifies
whether the attacker must engage in multiple rounds or
\textit{continuous} participation to achieve success or can
accomplish the attack in a single, \textit{one-shot} instance}. 

\hl{%
This survey categorizes attacks based on their specific nature and
tactics, offering a detailed taxonomy and examining their impacts on FL
systems. To provide a structured overview, we present a comprehensive
overview in Table}~\ref{tab:attack_categories}\hl{ summarizing various
attack types and categorizing them based on the mentioned dimensions.
Although we define five primary dimensions--target specificity, phase
affected, intent, nature of the adversary, and
execution style--real-world attacks can exhibit traits spanning more than
one category. For instance, a poisoning attack might initially appear
\emph{untargeted} but also target a specific class or region of the
data. Likewise, an \emph{insider} adversary could collaborate with
\emph{outsider} entities or extend the attack from training into
inference phases. In Table}~\ref{tab:attack_categories}\hl{, we classify
each attack according to its most typical or principal form while
recognizing that adversaries can mix methods or adopt hybrid
strategies.} The following sections will discuss critical security
threats in FL, detailing their nature, objectives, and potential impacts
and providing examples from the literature.

\begin{figure}[!ht]
  \centering
  \includegraphics[width=\linewidth]{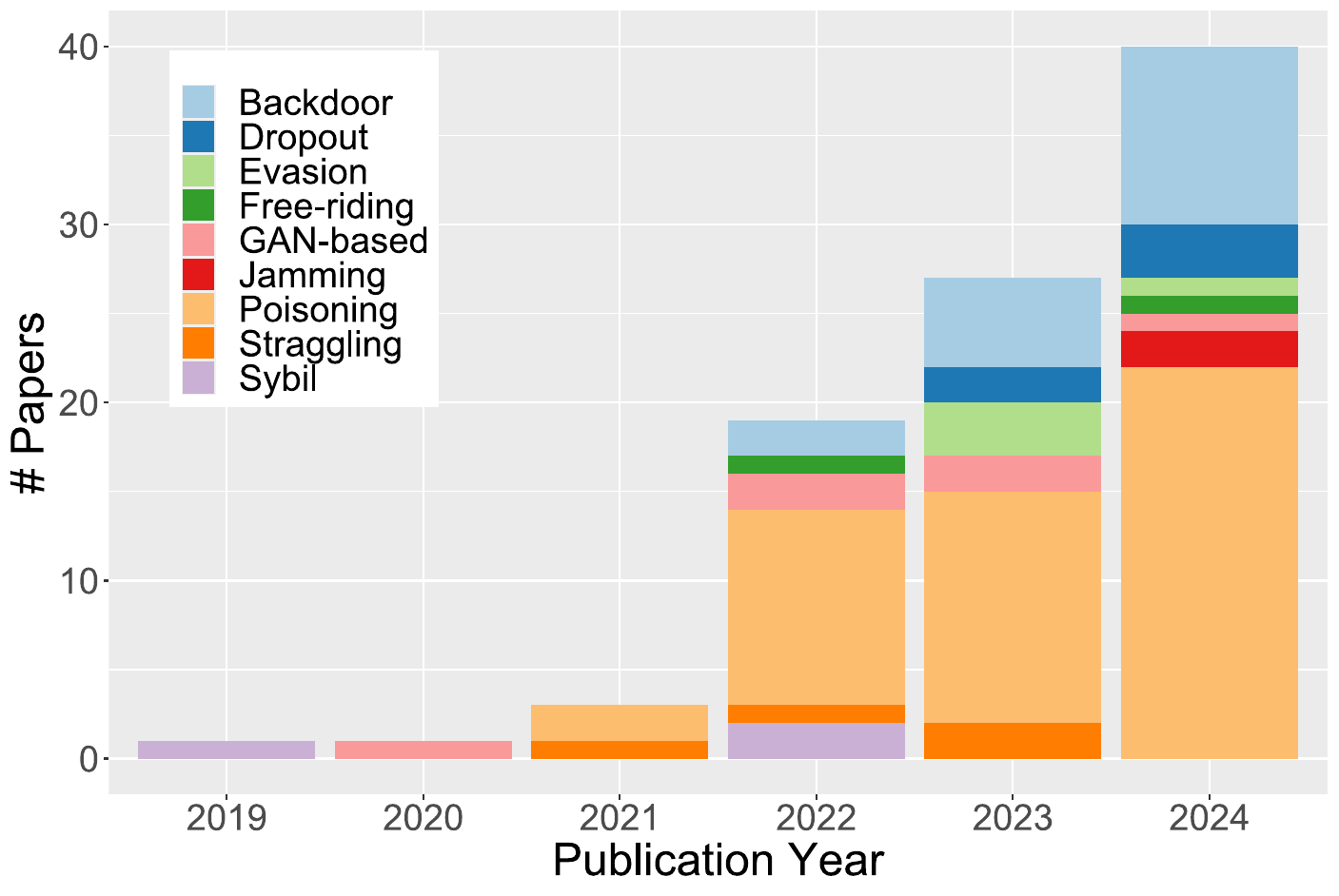}
  \caption{Papers related to security attacks over time}
  \label{fig:security_attacks_in_time}
\end{figure}

Fig.~\ref{fig:security_attacks_in_time} shows a clear upward trend in
the number of papers published on various security attacks over time. In
2019 and 2020, very few papers focused on Sybil and GAN-based attacks,
respectively. From 2021 onwards, there's a noticeable diversification in
the types of attacks studied, with a significant increase in overall
research output. Poisoning attacks have become increasingly prominent,
dominating the research landscape, especially in 2023 and 2024.  Other
attack types like backdoor, dropout, evasion, and free-riding have
emerged in the later years, indicating an expansion in the scope of
security research. 2024 shows the highest number of papers across
multiple attack categories, suggesting a growing interest and concern in
security attacks.

\begin{table*}[!ht]
\centering
\caption{Categorization of Security Attacks in FL}
\resizebox{\textwidth}{!}{
\begin{tabular}{@{}llllll@{}}
\toprule
\textbf{Attack Type}    & \textbf{Target Specificity} & \textbf{Phase Affected}    & \textbf{Intent}         & \textbf{Nature of Adversary} & \textbf{Execution Style} \\ \midrule
\textit{Data Poisoning}  & Targeted/Untargeted      & Training                   & Malicious                & Insider/Outsider                     & Continuous \\
\textit{Model Poisoning} & Targeted/Untargeted                    & Training                   & Malicious                & Insider                     & Continuous \\
\textit{GAN-based}       & Targeted/Untargeted                    & Training                   & Malicious                & Outsider                     & Continuous \\
\textit{Sybil}           & Targeted/Untargeted                  & Training/Inference         & Malicious                & Insider/Outsider                     & Continuous \\
\textit{Backdoor}        & Targeted                    & Training                   & Malicious                & Insider                      & Continuous \\
\textit{Free Riding}     & Untargeted                  & Training                   & Exploitative             & Insider                      & Continuous \\
\textit{Jamming}         & Untargeted                  & Training/Inference         & Disruptive               & Outsider                     & One-Shot \\ 
\textit{Evasion}         & Targeted/Untargeted                  & Inference         & Disruptive               & Outsider                     & One-Shot \\ 
\textit{Straggling}      & Untargeted                  & Training/Inference         & Disruptive               & Insider             & Continuous \\
\textit{Dropout}         & Untargeted                  & Training/Inference         & Disruptive               & Insider             & Continuous \\

\bottomrule
\end{tabular}
}
\label{tab:attack_categories}
\end{table*}

\subsubsection{Byzantine attacks} 
\hl{%
A Byzantine attack refers to a broad category of malicious or faulty
behaviors within distributed and FL systems. The term originates from
the Byzantine Generals Problem}~\cite{lamport2019byzantine}\hl{, which
highlights the challenge of achieving consensus in a distributed network
when some clients act unpredictably due to malice or faults. In FL,
these attacks disrupt the learning process, degrade model performance,
or compromise system integrity. For instance, \emph{model and data
poisoning} attacks involve adversaries injecting harmful updates to skew
the global model, as shown in empirical studies where such attacks
significantly increase error rates}~\cite{fang2020local}\hl{. Similarly,
\emph{Sybil attacks} manipulate aggregation by introducing multiple fake
identities, amplifying the attacker's
influence}~\cite{shejwalkar2021optimizing}\hl{.  \emph{Backdoor
attacks}, on the other hand, secretly alter model behavior for
specific inputs, such as embedding triggers that activate malicious
outcomes. Real-world examples include tampering with IoT device models
to misclassify security threats or injecting biased data in healthcare
applications to compromise diagnostic accuracy.} \hl{In practical terms,
adversaries can perform Byzantine behaviors by intercepting local
gradient updates and introducing arbitrary deviations before sending
them to the server.  Minimal Python scripts can scale or randomize these
updates, allowing the attacker to bypass naive filters. Some open-source
prototypes demonstrate how two or three malicious clients can
systematically skew the global model.  Moreover, robust aggregation
methods (e.g., Bulyan, Krum) typically detect large outliers but may
fail against subtle manipulations. Integrating cryptographic checks
(e.g., commitments) or analyzing multi-round consistency across updates
can significantly reduce the success rate of Byzantine exploits.}
\hl{The following paragraphs explore these attack mechanisms and their
consequences in detail.}

\paragraph{Poisoning attacks}
Poisoning attacks in FL involve injecting malicious data or manipulating
model updates to compromise the integrity of the learning process. Such attacks can decrease overall model performance and allow the attacker to
introduce biases, insert backdoors, or create specific targeted
vulnerabilities. 

\hl{%
\textbf{Data poisoning} refers to attacks where malicious clients
alter their data or model's parameters sent to the global model to
degrade its performance. \textit{Untargeted data poisoning} involves
general disruptions, such as adding random noise, random label flipping,
and random input data poisoning, which can cause a significant drop in
model accuracy and robustness. For instance, an attacker injecting noisy
data into medical diagnosis models can lead to incorrect patient
assessments. \textit{Targeted data poisoning}, on the other hand, seeks
to cause specific errors or misclassifications, such as targeted label
flipping in autonomous driving systems, where stop signs are
misclassified as speed limits, posing safety risks. Another way to
categorize data poisoning attacks is based on whether the attacker can
modify the labels of the poisoned data. \textit{Clean-label poisoning}
assumes that attackers cannot change data labels due to integrity
constraints but instead subtly manipulate features, such as modifying
image pixels to induce incorrect classifications. These attacks are
especially dangerous in security-sensitive domains like biometric
authentication, where small perturbations in face recognition models can
allow unauthorized access while remaining undetectable by traditional
defenses.}

In contrast, \textit{dirty-label poisoning} involves directly
manipulating the data labels. In this scenario, the attacker introduces
samples into the training dataset with incorrect labels, misleading the
model during training. Dirty-label poisoning is generally easier to
detect than clean-label poisoning because the data and its labels'
inconsistencies are more apparent~\cite{nguyen2024wicked}.
\autoref{tab:poisoning_attacks} provides an overview of data poisoning
attacks, categorized by their targeted or untargeted nature and whether
they involve clean or dirty labels. The table includes relevant
papers for each category illustrating key research and findings.

\begin{table}[h!]
\centering
\caption{Classification of Data Poisoning Attacks}
\begin{tabular}{p{1.5cm} p{2.75cm} p{2.75cm}}
\hline
\textbf{} & \textbf{Clean-label} & \textbf{Dirty-label} \\
\hline
\textbf{Targeted} & Targeted data manipulation~\cite{shafahi2018poison}~\cite{nguyen2024wicked}~\cite{zhang2021poisongan}  & Targeted label flipping~\cite{gupta2023novel}~\cite{tolpegin2020data}~\cite{zhang2021poisongan}~\cite{sun2024attacking} \\

\textbf{Untargeted} & Random data manipulation~\cite{zhang2021poisongan}, Adversarial samples~\cite{shi2022data} & Random label-flipping~\cite{shahid2022label}~\cite{zhang2021poisongan} \\ 
\hline
\end{tabular}
\label{tab:poisoning_attacks}
\end{table}

\hl{%
\textbf{Model poisoning} involves the deliberate manipulation of model
parameters or updates sent to a central server, targeting the integrity
of the model itself rather than the training data. This type of attack
is particularly effective, often surpassing data poisoning in impact,
especially against systems employing Byzantine-robust defense
mechanisms. Fang et al.}~\cite{fang2020local}\hl{ demonstrated that
non-directional attacks, which craft local model parameters to deviate
significantly from expected values, can lead to aggregated updates that
degrade global model performance. For example, their experiments showed
that introducing perturbations maximized deviation from the typical
update path, resulting in substantial global model errors. Baruch et
al.}~\cite{baruch2019little}
\hl{highlighted that even minimal poisoning--where only a small fraction
of malicious updates is introduced--can bypass robust defenses by
exploiting gradient variance. This approach requires limited knowledge
of client data and subtly shifts the mean of aggregated gradients
to evade detection.  Real-world implications include attacks on
recommendation systems, where subtle manipulations degrade ranking
accuracy without triggering alarms. Wang et
al.}~\cite{wang2023analysis}\hl{ extended these findings to federated
online learning to rank (FOLTR) systems, showing that sophisticated
poisoning strategies outperform data poisoning even under robust
defenses. They also noted that deploying such defenses without active
attacks can degrade system performance, underscoring the need for
adaptive defenses that balance security and functionality.}

\hl{%
Implementation-wise, data-poisoning attacks often involve
straightforward label manipulation or pixel-level perturbations in the
local dataset. Publicly available code, such as
in}~\cite{fang2020local}\hl{, shows how a simple gradient-scaling
procedure can overpower benign updates in an aggregation function like
FedAvg. Model poisoning goes a step further, directly adjusting weight
tensors to embed "invisible triggers". Defenders typically integrate
robust aggregator pipelines (e.g., Krum or Trimmed Mean, see
Section}~\ref{subsec:securitydefenses})\hl{ and anomaly monitors
that track suspicious gradient magnitudes
or label discrepancies across rounds. Additionally, partial
local data checks (for example, removing highly implausible
labels) can disrupt stealthy poisoning attempts before
they aggregate into a global parameter shift.}

\paragraph{Generative Adversarial Network-based (GAN) attacks}
\hl{GANs have been employed to execute both model and data poisoning
attacks in FL. In such scenarios, an adversary masquerades as a benign
client and trains a GAN to replicate prototypical samples from
other clients' datasets. The global model parameters serve as the
discriminator's parameters, enabling the GAN to produce realistic yet
manipulated samples. These samples are then used to generate poisoning
updates, which are scaled and submitted to the central
server}~\cite{qiao2023privacy}\hl{. According to Zhang et
al.}~\cite{zhang2019poisoning}\hl{, any internal client can
initiate GAN-based poisoning attacks. For instance, their PoisonGAN
model demonstrated that even under attack, the global model retained
over 80\% accuracy on both poisoning and primary
tasks}~\cite{zhang2021poisongan}. \hl{This highlights the dual threat of
maintaining task performance while embedding malicious objectives.
Real-world implications include adversaries exploiting GANs to bypass
detection mechanisms, as seen in cases where vague or noisy poisoned
data undermines anomaly detection systems}~\cite{sun2024gan}\hl{. These
examples underline the significant harm GAN-based attacks pose, which
compromise FL systems' integrity and privacy without easily detectable
anomalies.} \hl{From the implementation perspective, a GAN-based attack
typically involves pairing the server's global model (as a
discriminator) with a locally trained generator that refines malicious
updates to appear ``benign.'' Minimal modifications to PyTorch or
TensorFlow scripts let attackers pass generator outputs as legitimate
gradients. Potential defenses could include incremental offset detection
that flags suspiciously consistent gradient distortions and clustering
techniques for client updates with significant divergences.}

\paragraph{Sybil attacks}
\hl{%
The Sybil attack involves a malicious client creating multiple fake
identities to gain disproportionate influence or control over the
system. While not specifically a poisoning attack, it can facilitate or
amplify poisoning attacks by increasing the number of fake clients that
submit malicious or biased updates}~\cite{neto2023survey}\hl{.  For
example, model poisoning attacks using fake clients can significantly
reduce the test accuracy of the global model, even against classical
defenses}~\cite{cao2022mpaf}\hl{. Fung et al.}~\cite{fung2020mitigating}
\hl{demonstrated this in their experiment where two Sybil nodes inserted
a backdoor, causing 96.2\% of digit 1s in the MNIST dataset to be
misclassified as 7s in the final model. This highlights the severe
impact even a few Sybils can have on model integrity. In real-world
scenarios, such attacks are particularly concerning due to their ability
to bypass detection mechanisms by preserving overall model
utility}~\cite{fung2020mitigating}\hl{.  Furthermore, another
study}~\cite{xiao2022sbpa}\hl{ revealed how Sybil nodes could inject
backdoor triggers into data, disrupting training processes in FL
systems.}

\hl{Employing Sybil clients can be as simple as registering multiple
``fake'' clients that communicate identical or slightly modified
updates, all controlled by one adversary. Code
examples}~\cite{fung2020mitigating}\hl{ illustrates that only two Sybils
can drastically corrupt a federated model.
FoolsGold}~\cite{fung2020mitigating}\hl{, or other similarity-based
approaches track the cosine distance among client updates, penalizing
suspicious clusters. Some frameworks incorporate blockchain-based
identity management or limit how many new clients can join per
round, raising barriers for mass Sybil infiltration. These
countermeasures reduce the effectiveness and stealth of Sybil-based
manipulations.}

\paragraph{Backdoor attacks}
\hl{Backdoor attacks are a form of targeted poisoning attack in which an adversary deliberately corrupts the global model, making it perform well on the main task while exhibiting malicious behavior when triggered by specific conditions, such as a particular label, image modification, or feature}~\cite{mothukuri2021survey}\hl{. These attacks are particularly concerning in FL due to the decentralized nature of training, where malicious updates can propagate vulnerabilities across the entire system. For example, in real-world scenarios like next-word prediction models used in mobile applications, backdoor triggers could manipulate outputs for sensitive contexts, such as political events}~\cite{liu2023facilitating}\hl{. Liu et al.}~\cite{liu2023facilitating}\hl{ demonstrated that backdoor attacks could accelerate FL convergence by crafting local updates that mimic global data distributions and injecting backdoors during later stages when benign updates have minimal impact. However, these attacks face challenges such as detection risks and limited persistence. Dai et al.}~\cite{dai2023chameleon}\hl{ addressed these issues by proposing the Chameleon attack, which uses poisoned datasets and contrastive learning to enhance backdoor durability. This method ensures the backdoor remains effective even after attackers stop participating, as seen in applications like IoT devices with weak security measures}~\cite{dai2023chameleon}\hl{. Similarly, Zhang et al.}~\cite{zhang2023a3fl}\hl{ highlighted that fixed backdoor triggers often fail under global training dynamics. Their A3FL approach adapts triggers adversarially to maintain effectiveness in evolving models. These examples underscore the significant harm of backdoor attacks, which can compromise model integrity and user trust in critical applications like autonomous vehicles or healthcare systems.} \hl{Implementing a backdoor often involves a "trigger pattern" integrated into a small fraction of the local training set (e.g., a tiny corner pixel pattern in image classification). Attack scripts typically swap labels for these trigger-laden inputs and train locally to ensure the global model learns to misclassify only when the pattern appears. FLAME}~\cite{nguyen2022flame}\hl{ and other advanced defenses add mild noise or rely on "clean validation" to detect unexpected performance spikes on specific triggers. Another method is partial neuron pruning, removing neurons that show abnormally high activation for certain triggers. Adopting these defenses usually increases training overhead but significantly reduces successful backdoor injection rates.}

\subsubsection{Free-Riding}
\hl{Free-riding occurs when a client benefits from the final aggregated model without contributing to its training due to reasons such as lack of data, privacy concerns, or insufficient computational resources. In the context of Free-Rider (FR) attacks, these can be categorized into Anonymous Free-Rider (AFR) and Selfish Free-Rider (SFR) attacks based on the adversary's control over private data and computing resources}~\cite{wang2024pass}\hl{. AFR attackers, lacking private datasets or computational resources, typically contribute stochastic Gaussian noise to the central server, resembling a generic Gaussian attack}~\cite{fraboni2021free}\hl{. This behavior undermines model accuracy by introducing noise into the aggregation process. In contrast, SFR attackers possess private data and computational abilities but choose not to contribute these resources. For instance, SFR attackers may employ advanced strategies like delta weights attacks, generating gradient updates by subtracting two global models from previous rounds}~\cite{lin2019free}\hl{, or submit systematically crafted fake parameters}~\cite{domingo2022secure}\hl{. While delta weights attacks ensure convergence of the aggregated model, they maintain stealth by mimicking benign updates}~\cite{zhu2021advanced}\hl{. Even simpler methods, such as consistently returning the same global model parameters, can degrade model performance and reduce fairness in FL}~\cite{mothukuri2021survey}\hl{. These attacks pose significant threats in real-world scenarios, especially in sensitive domains like healthcare or finance, where FL's integrity is crucial}~\cite{fraboni2021free}. \hl{From an implementation perspective, a free-rider can bypass local training entirely by returning either unchanged or random parameters while continuing to download global updates. These minimal modifications exploit the aggregator's inherent trust in each client. PASS}~\cite{wang2024pass}\hl{ and similar auditing approaches evaluate each client's historical gradient contributions against their impact on model improvements. Clients that fail to provide meaningful updates risk detection or a reduced aggregation weight. These scoring mechanisms discourage free-riders by linking model benefits to local effort.}

\subsubsection{Jamming Attacks}
Jamming attacks pose a severe security threat in wireless networks, particularly decentralized FL (DFL) environments~\cite{beltran2023decentralized}. These attacks involve adversaries emitting interference signals to disrupt communication between legitimate nodes, hindering the exchange of critical data such as local model parameters. For example, in real-world scenarios like airport operations, jamming has led to significant disruptions in communication systems, delaying processes and compromising operational efficiency~\cite{shi2023jamming}. In blockchain-based decentralized FL, jamming attacks prevent normal miners from receiving necessary data, excluding them from proof-of-work computations. This gives malicious miners an advantage in controlling the blockchain by increasing the probability of generating a longer malicious block stream, especially when the number of attackers surpasses normal miners~\cite{kim2024thethreat}. Additionally, targeted jamming in decentralized FL can isolate nodes by disrupting key communication links. This isolation fragments the network, delaying learning processes and degrading model accuracy due to insufficient data exchange~\cite{shi2023jamming}. For instance, simulations of such attacks on multi-hop wireless networks have demonstrated significant reductions in DFL performance by exploiting vulnerabilities in connectivity and model sharing~\cite{shi2023jamming}. Realistic jamming can be emulated by imposing network drop rates or forced timeouts in each training round. Indeed, attackers might saturate specific channels, delaying or preventing the arrival of local updates to the server. From the defense perspective, coded computations (e.g., CodedPaddedFL~\cite{schlegel2022codedpaddedfl}) or asynchronous protocols allow partial aggregation even if a subset of updates is lost or late. Additionally, some FL systems introduce fallback communication channels to bypass jammed links. These solutions provide a certain level of robustness to partial network disruption.

\subsubsection{Evasion Attacks}

\hl{Evasion attacks exploit weaknesses in model predictions during inference by introducing carefully crafted adversarial inputs, such as pixel perturbations, without altering the training process}~\cite{kumar2024revamping}\hl{. For instance, unnoticeable changes to a panda image can cause GoogLeNet to misclassify it as a gibbon with 99.3\% confidence}~\cite{li2024threats}\hl{. These attacks undermine the reliability of FL systems by reducing model accuracy and trustworthiness. In real-world scenarios, evasion attacks can deceive spam filters or recommendation systems trained via FL, leading to financial or operational harm}~\cite{kim2023pfeddef}\hl{.

In VFL, Pang et al.}~\cite{pang2022adi}\hl{ demonstrate the susceptibility of VFL systems to ADIs, which manipulate joint inference outcomes to prioritize an attacker's input. They employ gradient-based methods and grey-box fuzz testing to uncover vulnerabilities in privacy-preserving features, revealing that adversaries can exploit these to skew results. For example, ADIs could be used in financial applications to favor fraudulent transactions. To address these threats, Kim et al.}~\cite{kim2023pfeddef}\hl{ analyze internal evasion attacks across learning methods, showing that personalized federated adversarial training enhances robustness by 60\% compared to standard approaches. This demonstrates that tailored defenses can mitigate attack impacts even under constrained resources, though challenges remain in balancing accuracy and security.} \hl{To carry out such an attack, malicious entities could leverage adversarial example implementations to craft feature-level perturbations. Only minor changes to the inference pipeline could be enough to cause misclassifications in the global model. To mitigate the impact, using personalized adversarial training} ~\cite{kim2023pfeddef}\hl{ allows for retraining on adversarial variants each round, though at a higher computational cost.}

\subsubsection{Straggling}
\hl{Sometimes, due to various factors like limited computing resources, background processes, memory constraints, or unstable wireless communication, certain edge devices, known as stragglers, might perform significantly slower than others, thereby deteriorating the FL process. This vulnerability can also be exploited by adversaries through free-riding attacks, where malicious clients intentionally delay or avoid computations to degrade system performance}~\cite{li2023fedvs}\hl{. Waiting for model updates from these slower clients at each learning step can slow down model convergence and degrade accuracy. For instance, attackers may inject noise into updates or mimic benign clients to amplify delays, leading to inefficient resource utilization as faster clients idle}~\cite{park2021sageflow}\hl{. Ignoring updates from stragglers risks model accuracy and client drift -- a phenomenon where local models diverge significantly due to non-identically distributed data. Real-world manifestations include healthcare FL systems where malicious clients disrupt timely updates, jeopardizing critical applications like disease prediction.} \hl{In terms of implementation, simple modifications in local training scripts can pause or throttle GPU usage, slowing progress. Therefore, the design of asynchronous or coded protocols is required to reduce reliance on a strict round barrier. If certain clients are repeatedly late or absent, they can be down-weighted or removed from the aggregator's pipeline. Nonetheless, balancing the fair inclusion of actual slow clients against malicious stragglers remains a key design challenge in practical FL settings.}

\subsubsection{Dropout} \hl{User dropout in FL refers to the scenario in which some clients drop out or become inactive during training. This phenomenon can occur due to network issues, client failures, or intentional withdrawal. Honest clients may become demotivated to engage in the training process if the collaborative framework does not guarantee fairness for all clients}~\cite{chowdhury2023fedrlchain}\hl{. Beyond these general challenges, dropout can also manifest as an attack, where malicious clients intentionally withdraw at critical training stages to disrupt the global model's convergence. Such targeted dropout attacks can exacerbate biases in the model if specific clients with unique data distributions are affected, leading to skewed performance}~\cite{schlegel2022codedpaddedfl}\hl{. For instance, in real-world scenarios like healthcare applications, the dropout of clients representing minority populations could result in a poorly performing model on underrepresented groups.} \hl{In code, dropout simulates a failure to send updates by skipping the aggregator's communication calls. Defensive solutions require tracking dropout patterns over time to determine if certain clients drop out from training at crucial convergence stages. The integration of partial reweighting or client selection}~\cite{deressa2024trustbandit}\hl{ may reduce the damage, though guaranteeing fairness if many dropouts occur remains non-trivial.}

We would like to note that while straggling and dropout are not traditionally categorized as intentional attacks in FL (highlighted in gray on the taxonomy of Fig.~\ref{fig:taxonomy_sec_priv}), they represent significant challenges that can hinder the overall learning process.  However, it is important to note that these phenomena could also be exploited by adversaries in a malicious context. An attacker could deliberately induce straggling by compromising clients or resources or cause dropout by intentionally withdrawing specific clients to disrupt the training process.

\subsection{Security Defenses}
\label{subsec:securitydefenses}
\hl{This section provides an overview of security mechanisms designed to enhance the robustness of FL systems against various adversarial threats. It highlights key strategies, including robust aggregation operators, anomaly detection techniques, and adversarial training.}

\subsubsection{Robust Aggregation Operators}

FedAvg is one of the most popular algorithms used in FL to aggregate client model updates. However, several studies have shown that this method can be sensitive to various types of attacks, including model poisoning attacks, where some clients might send malicious updates, or data poisoning attacks, where the data used to train local models is manipulated to bias the global model~\cite{blanchard2017byzantine}~\cite{xie2018generalized}. Robust aggregation operators have been developed to enhance security and defend against such attacks. These operators are designed to minimize the impact of malicious or noisy updates, thereby improving the resilience of the FL system.

\begin{itemize}
    \item \textbf{Trimmed Mean} involves calculating the average of model updates after removing a specified percentage of the highest and lowest values. This method helps mitigate the impact of outliers but can be circumvented by poisoning attacks that exploit high empirical variance among client updates, as demonstrated by "A Little Is Enough" ~\cite{Li2024experimental}. \hl{This solution also mitigates the reduction in performance caused by non-IID data by removing extreme values from clients whose distributions differ significantly from the rest.}

    \item \textbf{Median}-based algorithms replace the arithmetic mean with the median of model updates, choosing the value representing the distribution's center. This approach is less sensitive to extreme values and more resistant to adversarial attacks compared to methods like FedAvg. \hl{This approach also improves the model performance under high non-IID data since it aims to avoid the influence of highly different distributions (a.k.a outliers)}. However, it is vulnerable to attacks such as IPM, which can negatively impact the inner product between the true gradient and the aggregated gradients~\cite{Li2024experimental}. \textbf{GeoMed (Geometric Median)} minimizes the sum of Euclidean distances to all points, offering a central point that is less sensitive to outliers compared to the mean~\cite{pillutla2022robust}. Its more computation-efficient version is called Medoid~\cite{xie2018generalized}. GeoMed can tolerate up to half of the malicious clients and estimate true parameters, showing convergence properties in gradient descent methods. However, GeoMed is sensitive to model poisoning attacks and less robust with imbalanced datasets. To address these issues, Li et al.~\cite{li2023byzantine} proposed \textbf{Auto-Weighted GeoMed (AutoGM)}, which automatically excludes extreme outliers and re-weights remaining points based on a user-specified skewness threshold. AutoGM maintains high performance even with up to 30\% of nodes engaging in model poisoning or 50\% experiencing data poisoning attacks. \textbf{Marginal Median (MarMed)}~\cite{xie2018generalized} focuses on the median of marginal distributions of data points, filtering out extreme values to provide a stable estimate of central tendency. This approach, similar in robustness to the geometric median but with a distinct handling of data, helps maintain the integrity of the aggregation process against adversarial manipulations. \textbf{Mean Around Median (MeaMed)}~\cite{xie2018generalized} is a trimmed average method that centers calculations around the median, effectively reducing the impact of outliers and adversarial data. Blending the strengths of both the mean and median offers a balanced approach to maintaining performance and robustness in distributed learning scenarios vulnerable to Byzantine attacks.

    \item \textbf{Krum}, introduced by Blanchard et al.~\cite{blanchard2017byzantine}, selects a model update vector that is least affected by outliers by minimizing the sum of squared distances to its \(n-f\) closest neighbors, where \(f\) is the maximum number of Byzantine workers tolerated. \textbf{Multi-Krum} (or m-Krum) extends this approach by considering multiple vectors, thus enhancing robustness by aggregating \(d\) parameter vectors instead of just one. Despite its effectiveness in mitigating high-severity attacks, Han et al.~\cite{han2024fedsecurity} found that Krum struggles with RNNs due to variability in local models caused by sequential data and recurrent structures. Additionally, Krum's reliance on strong assumptions, such as bounded absolute skewness, may not always be realistic, and it is vulnerable to newer attacks like IPM and "A Little Is Enough" (ALIE), which exploit empirical variances between client updates~\cite{Li2024experimental}.

    
    \item \textbf{Bulyan} enhances existing Byzantine-robust aggregation techniques, such as Krum and GeoMed, by first compressing gradient updates from each client into a more compact form. This reduces the impact of noise and malicious data. After compression, Bulyan employs a robust aggregation technique to combine the compressed updates, focusing on reliable information while filtering out outliers and adversarial contributions, thereby improving accuracy and resilience against Byzantine faults~\cite{mhamdi2018hidden}.
    
    \item \textbf{Clustering} aggregation calculates pairwise cosine distances between parameter updates and groups clients based on cosine similarities using agglomerative clustering with average linkage. While this method shows robustness in some scenarios, it only considers the relative directions of updates, ignoring their magnitudes. Attackers can exploit this by amplifying their updates without altering directions, disrupting model convergence. To address this, Li et al.~\cite{Li2024experimental} proposed \textbf{ClippedClustering}, which applies norm-based clipping to updates. Updates are scaled if their norm exceeds a server-determined threshold, set automatically using the median of historical update norms, improving defenses under IID local datasets. However, ClippedClustering significantly degrades performance with non-IID datasets, highlighting the need for tailored defense strategies.

    \item \textbf{Zeno}~\cite{xie2019zeno} scores and ranks updates based on their alignment with a reference gradient, filtering out suspicious updates dynamically. Zeno is particularly effective in resisting Byzantine attacks because it relies not solely on traditional statistical measures like medians or means. Instead, it actively evaluates the credibility of each update, allowing it to reject harmful contributions dynamically. In contrast to previous work, Zeno++~\cite{xie2021zeno} removes several unrealistic restrictions on worker-server communication, now allowing for fully asynchronous updates from anonymous workers, for arbitrarily stale worker updates, and for the possibility of an unbounded number of Byzantine workers. 
    
    \item \textbf{Anomaly Detection} It employs various statistical and analytical methods to identify events that deviate from expected behavior, which is crucial for detecting Byzantine attacks. Effective anomaly detection systems require a normal behavior profile to recognize malicious activity. Techniques might include clustering to group similar updates and identify outliers, Euclidean distance metrics used in methods like Krum for detecting deviations in input parameters, Autoencoders that reconstruct data to flag abnormal updates, and other methods~\cite{mothukuri2021survey}. For example, Jiang et al.~\cite{jiang2021sybil} proposed monitoring the average loss reported by clients to identify and exclude potentially compromised updates caused by Sybil attacks. Pan et al.~\cite{pan2024oneshot} proposed integrating advanced anomaly detection techniques with a unique model update aggregation strategy, enabling the identification and neutralization of backdoor influences in a single update cycle, avoiding the need for extensive data access or communication between clients. \hl{Since non-IID data can make normal client updates appear like anomalies, which attackers may exploit, adaptive anomaly detection methods such as the one proposed by Jiang et al.}~\cite{jiang2021sybil}\hl{ and Pan et al.}~\cite{pan2024oneshot} \hl{help to differentiate between natural variations in data and adversarial manipulations.}
\end{itemize}

\subsubsection{Asynchronous Schemes}
To mitigate the straggler issue, various asynchronous schemes have been proposed. These schemes update the global model based on the time difference between the current round and the previous round when the client first received the global model~\cite{park2021sageflow}. For example, Lu et al.~\cite{lu2024adaptive} proposed FedAAM, which employs an adaptive weight allocation algorithm that assigns dynamic weights to client updates based on their contribution, considering factors such as the timeliness and quality of the updates. The framework introduces two asynchronous global update rules based on a differentiated strategy, allowing the global model to update with varying client contributions depending on their performance and the frequency of their updates. Additionally, FedAAM integrates global momentum by using the historical global update direction, which helps smooth the global update process and manages the asynchrony among clients, thereby improving training efficiency and convergence behavior. However, Schlegel et al.~\cite{schlegel2022codedpaddedfl} report that these schemes generally do not converge to the global optimum. They further propose two schemes to avoid this problem. CodedPaddedFL combines one-time padding with gradient codes to ensure straggler resiliency while maintaining privacy, achieving an 18x speed-up for 95\% accuracy on the MNIST dataset. CodedSecAgg, based on Shamir's secret sharing, provides both straggler resiliency and robustness against model inversion attacks, outperforming the state-of-the-art LightSecAgg by a speed-up factor of 6.6-18.7 for similar accuracy.

\subsubsection{Pruning} Pruning can serve as both an optimization strategy and a potential security measure in FL. It is a technique used to reduce the size and complexity of machine learning models by removing less important or dormant neurons and connections. This process helps address the computational and communication constraints typical in FL environments, where clients often have limited resources~\cite{mothukuri2021survey}. Additionally, selective pruning can enhance security and mitigate backdoor attacks by removing neurons that are not activated by clean data. However, this defense method may be less effective if attackers use pruning-aware methods~\cite{liu2023facilitating}.

\subsubsection{Adversarial Training} 
Adversarial training in FL is a defense mechanism in which each client generates adversarial examples locally during training to enhance the robustness of their model updates against adversarial attacks. These adversarially trained local models are then aggregated by the central server, allowing the global model to learn to resist adversarial inputs without directly exposing client data, thereby improving security in a distributed and privacy-preserving manner. Li et al.~\cite{li2023federated} formulated the training process as a min-max optimization problem, addressing the unique challenges of decentralized data and model training. They also provided a detailed convergence analysis, demonstrating that the minimum loss can converge to a small value under appropriate conditions, and introduced gradient approximation techniques to enhance training effectiveness, particularly for non-IID clients.

\subsubsection{Personalized Solutions}
This section covers customized solutions that do not fit the categories discussed above. It highlights the most novel and promising methods from recent research, showcasing innovative approaches to enhancing security in FL environments.

\emph{FoolsGold} It is a defense method specifically designed to counter targeted poisoning sybil attacks. It identifies clients with similar behavior and characteristics of Sybil clones. It then adapts the learning rates of these clients based on the similarity of their contributions, effectively reducing the influence of malicious updates and mitigating the attack~\cite{fung2020mitigating}~\cite{fung2020limitations}. This technique also provides a way to tackle non-IID data issues presented in Section~\ref{sec:non-iid-impact}, detecting overrepresented gradients and down-weighting contributions from clients that exhibit unusually high similarity, ensuring fairer aggregation despite heterogeneous data variations. However, when legitimate updates are similar, these methods also tend to penalize them, causing significant drops in model performance~\cite{jebreel2022fldefender}. Some experiments have shown that FoolsGold might completely fail to train the model, potentially eliminating important local models. Additionally, the method may encounter limitations when integrated with large language models due to the substantial cache requirements needed to memorize intermediate results, such as models from previous FL training rounds~\cite{han2024fedsecurity}.

\emph{FL-Defender}~\cite{jebreel2022fldefender}: Proposed by Jebreel et al., FL-Defender is a defense mechanism designed to combat targeted poisoning attacks, specifically addressing label-flipping (a type of poisoning attack) and backdoor attacks. Similarly to FoolsGold, FL-Defender extracts last-layer gradients from workers' updates and calculates cosine similarities to detect attack patterns, followed by dimensionality reduction using PCA to focus on the most relevant features. During aggregation, it re-weights updates based on historical deviations, minimizing the influence of poisoned data while preserving model performance and maintaining low computational overhead. Its aggregation method is similar to Krum and Trimmed Mean. However, FL-Defender adds an adaptive component by re-weighting updates rather than rejecting them outright.

\emph{FLAME}~\cite{nguyen2022flame}: It is a robust defense framework designed to counter backdoor attacks while preserving the benign performance of the global model, even in non-IID data settings. Unlike traditional defenses that rely on limited attacker models or degrade performance with excessive noise, FLAME dynamically estimates and injects the optimal amount of Gaussian noise to eliminate backdoors. Its clustering-based approach effectively separates malicious updates from benign ones, ensuring robust aggregation despite data heterogeneity. Additionally, weight clipping limits the influence of outlier updates, enabling FLAME to maintain high model accuracy while efficiently removing adversarial backdoors. 

\emph{PAMPAS}~\cite{ching2020model}: Ching et al. proposed to combat GAN attacks by partitioning the model between users and edge servers, with users training only part of the model to enhance security and efficiency. Their approach seeks to optimize model partitioning to resist GAN attacks and minimize total training time while addressing the trade-offs between computation, transmission, and maintaining data privacy.

\emph{PPFDL}~\cite{xu2022privacy}: Xu et al. proposed a solution designed to reduce the negative impact of irregular users (Users who join and leave the training process frequently or unpredictably) on training accuracy by prioritizing high-quality data contributions. The approach ensures the confidentiality of user information using Yao's garbled circuits and additively homomorphic cryptosystems. PPFDL is also robust against user dropout, allowing the training to continue as long as some users remain online. 

\emph{LeadFL}~\cite{zhu2023leadfl}: Zhu et al. proposed a client-side defense mechanism against backdoor and poisoning attacks, which introduces a novel regularization term in local model training to nullify the Hessian matrix of local gradients. Additionally, the regularization helps to tackle non-IID issues explored in Section~\ref{sec:non-iid-impact} by neutralizing adversarial gradient patterns, improving robustness against backdoor and targeted attacks in heterogeneous data settings. Unlike existing defenses, LeadFL specifically targets the Hessian matrix to enhance robustness against bursty adversarial patterns, effectively handling the high variance in malicious client activity that many server-side defenses struggle with. Designed to work alongside existing server-side defenses, LeadFL enhances overall security by complementing other mechanisms rather than functioning as a standalone solution. 


\emph{PASS}~\cite{wang2024pass}: To address Free-Rider attacks in FL, the paper introduces the Parameter Audit-based Secure and Fair FL Scheme (PASS). PASS employs a privacy-preserving strategy (PASS-PPS) incorporating weak DP with a Gaussian mechanism and a parameter prune mechanism to protect data during parameter auditing. Additionally, PASS utilizes a novel contribution evaluation method to accurately measure each client's performance, ensuring fairness in the training process and deterring both AFR and Selfish SFR attacks.

\emph{Sageflow}~\cite{park2021sageflow}: It introduces a staleness-aware grouping method that integrates seamlessly with robust aggregation rules such as Multi-Krum. This approach enhances resilience against adversaries through entropy filtering and loss-weighted averaging, effectively managing non-IID data distributions and outperforming previous methods like Zeno+ in practical scenarios. 

\emph{FedRLChain}~\cite{chowdhury2023fedrlchain}: It leverages blockchain technology to address critical challenges in Federated Reinforcement Learning. This framework features a novel verification algorithm to counter malicious client actions, an aggregation weight scheme to avoid bias in the global model, and an enhanced FedAvg algorithm for improved convergence speed.

In Table~\ref{tab:attacks-defense-security-comparison}, we provide a relation between the defense mechanisms and the corresponding attacks or threats they aim to mitigate in FL. Certain defenses, like robust aggregation operators and anomaly detection, address various attacks such as poisoning, GAN-based, Sybil, backdoor, free-riding, jamming, and evasion. Asynchronous schemes, pruning, and personalized solutions focus more specifically on addressing straggling and dropout issues related to client heterogeneity and connectivity. 

\begin{table}[htbp]
    \centering
    \resizebox{\columnwidth}{!}{
    \begin{tabular}{@{}lccccccccc@{}}
          & \multicolumn{9}{c}{\textbf{Attacks/Threats}} \\
        \cmidrule{2-10}
        \textbf{Defenses}
        & \rotatebox{90}{\textbf{Poisoning}} 
        & \rotatebox{90}{\textbf{GAN-based}} 
        & \rotatebox{90}{\textbf{Sybil}} 
        & \rotatebox{90}{\textbf{Backdoor}} 
        & \rotatebox{90}{\textbf{Free-riding}} 
        & \rotatebox{90}{\textbf{Jamming}} 
        & \rotatebox{90}{\textbf{Evasion}} 
        & \rotatebox{90}{\textbf{Straggling}} 
        & \rotatebox{90}{\textbf{Dropout}} \\
        \midrule
        \textbf{Robust aggregation operators} & \textcolor{teal}{\ding{52}} & \textcolor{teal}{\ding{52}} & \textcolor{teal}{\ding{52}} & \textcolor{teal}{\ding{52}} & \textcolor{teal}{\ding{52}} & & \textcolor{teal}{\ding{52}} & & \\
        \textbf{Asynchronous schemes}         & \textcolor{teal}{\ding{52}} &     &     &     &     & \textcolor{teal}{\ding{52}} &     & \textcolor{teal}{\ding{52}} & \textcolor{teal}{\ding{52}} \\
        \textbf{Pruning}                      & \textcolor{teal}{\ding{52}} &     &     & \textcolor{teal}{\ding{52}} &     &     &     & \textcolor{teal}{\ding{52}} & \\
        \textbf{Adversarial training}         & \textcolor{teal}{\ding{52}} & \textcolor{teal}{\ding{52}} &     & \textcolor{teal}{\ding{52}} &     &     & \textcolor{teal}{\ding{52}} &     & \\
        \textbf{Personalized solutions}       &  \textcolor{teal}{\ding{52}}   &     &  \textcolor{teal}{\ding{52}}   &   \textcolor{teal}{\ding{52}}  &   \textcolor{teal}{\ding{52}}  &     &     & \textcolor{teal}{\ding{52}} & \textcolor{teal}{\ding{52}} \\
        \bottomrule
    \end{tabular}
    }
    \caption{Relationship of defense mechanisms and attacks for secure FL} 
    \label{tab:attacks-defense-security-comparison}
\end{table}

\hl{Thus, Table}~\ref{tab:attacks-defense-security-comparison}\hl{, together with the analysis of the papers assessed, reveals key insights into the current security landscape in FL by illustrating the effectiveness of various defense mechanisms against different types of attacks. A notable observation is the dominance of robust aggregation operators and anomaly detection techniques, which address the broadest range of threats. This suggests that adversarial manipulations, particularly poisoning, Sybil, and backdoor attacks, remain central concerns in FL security. However, these methods alone are not sufficient. For example, while robust aggregation enhances resilience against model and data poisoning, it does not directly counter jamming attacks, which disrupt communication rather than manipulate training data. 

Table}~\ref{tab:attacks-defense-security-comparison}\hl{ also highlights an ongoing challenge: no single defense mechanism can comprehensively mitigate all security threats, emphasizing the need for hybrid approaches. Personalized solutions and adversarial training offer promising advances by tailoring security mechanisms to specific attack vectors, but they remain underexplored in the context of free-riding and evasion attacks. The increasing sophistication of FL attacks necessitates continuous refinement of defense strategies, integrating multiple techniques to address emerging adversarial tactics holistically.}

\emph{Lessons learned:} The analysis of security attacks and defenses in FL reveals several critical insights. First, the growing sophistication of adversarial strategies highlights the need for adaptive and multi-layered defense mechanisms. While robust aggregation operators and anomaly detection remain foundational defenses, adversaries continuously develop novel poisoning and backdoor attack strategies that evade traditional filtering methods. \textcolor{black}{This aspect highlights the limitations of static, threshold-based defenses and motivates the need for more adaptive, real-time techniques.}
Second, client heterogeneity and participation dynamics significantly impact FL security, as attackers can exploit phenomena like straggling, dropout, and free-riding, emphasizing the importance of personalized and incentive-aligned solutions. \textcolor{black}{However, most current solutions assume honest or uniformly distributed clients, leaving a gap in defending against adversarial heterogeneity and collusion.}. Additionally, adversarial tactics such as Sybil-based collusion and GAN-powered attacks' rapid evolution underscores the necessity for continuous monitoring and adaptive countermeasures. \textcolor{black}{However, there is no standardized framework to evaluate these defenses across diverse threat types}. Finally, while many defenses focus on protecting global model integrity, there is a growing need for client-side security solutions to detect and mitigate threats locally before aggregation.

\subsection{Comparative Analysis of Defenses for Secure FL}

\hl{In this subsection, we analyze quantitative and experimental studies to evaluate the effectiveness of the previous defenses in specific scenarios. For instance, studies such as Li et al.}~\cite{li2023experimental}\hl{ and Zhang et al.}~\cite{zhang2025sok} \hl{offer a detailed quantitative comparison of various FL defense mechanisms across different attacks. Under untargeted model poisoning attacks like the Fang et al.}~\cite{fang2020local}\hl{ attack, Bulyan demonstrates superior robustness, achieving up to 85\% global model accuracy with a 20\% adversarial client ratio, compared to Trimmed Mean (78\%) and Krum (75\%). However, Bulyan's computational complexity ($O(dn^2)$ may hinder scalability in large-scale FL systems. For targeted backdoor attacks, FLTrust, which uses a small trusted dataset, achieves over 90\% accuracy on benign tasks while suppressing backdoor success rates below 5\%, outperforming Trimmed Mean, which achieves 85\% benign accuracy but struggles with backdoor suppression. FLAME emerges as a strong candidate in highly heterogeneous data settings by dynamically adds noise to mitigate backdoors while maintaining model performance at around 88\%  accuracy. Trimmed Mean balances simplicity and effectiveness for scenarios prioritizing low overhead. Thus, the choice of defense depends on the attack type and system constraints: Bulyan is recommended for untargeted attacks in smaller systems, while FLTrust and FLAME are preferred for targeted attacks or non-IID data distributions.}

\textcolor{black}{Beyond robust aggregation, the literature reports competitive results for the remaining four defense families in our taxonomy. \emph{Asynchronous schemes} such as FedAAM {\cite{lu2024adaptive}} and CodedPaddedFL {\cite{schlegel2022codedpaddedfl}} address stragglers and jamming by updating the global model as soon as partial gradients are available.  On MNIST, CodedPaddedFL provides around $95\%$ accuracy while delivering an $18\times$ reduction in time compared with synchronous FedAvg in settings with slow or jammed clients; the cost is roughly a two-fold increase in uplink bandwidth due to coded padding. Moreover, \emph{Pruning-based defenses} remove dormant or highly suspicious neurons after each aggregation round.  \cite{wu2020mitigating} shows that neuron pruning can cut a Fashion-MNIST backdoor attack success rate (ASR) from $99.7\%$ to $2\%$. \emph{Adversarial training} hardens the model against inference-time manipulations. For example, pFedDef~\cite{kim2023pfeddef} improves robustness PGD perturbations by roughly $60\%$ on CIFAR datasets while maintaining competitive clean accuracy. Finally, \emph{personalised solutions} mitigate Sybil and free-rider behaviour.  For example, Sageflow~\cite{park2021sageflow} further combines personalised weighting with entropy filtering, yielding a $12\%$ improvement in convergence speed under mixed Sybil-plus-straggler settings.}

These results confirm that each defense family excels under specific threat models and resource budgets: asynchronous protocols prioritise liveness, pruning targets stealthy backdoors, adversarial training bolsters prediction-time robustness, and personalised auditing enforces fairness against Sybil or free-riding behaviour.  A balanced deployment should therefore mix complementary mechanisms—for example, pairing Bulyan with FedAAM for integrity \emph{and} liveness, or coupling Trimmed-Mean with pFedDef to resist both poisoning and evasion—rather than relying on robust aggregation alone.

\section{Privacy in FL}
\label{sec:privacy}
\hl{Following the taxonomy depicted in Fig.}~\ref{fig:taxonomy_sec_priv}\hl{, we describe the main attacks and defenses for privacy in FL in this section. For the attacks, we provided specific mechanisms, degrees of harm, and specific examples of manifestations in the real world. At the end of each subsection, we provide some lessons learned after analyzing the papers regarding attacks and defenses for privacy in FL.}

\subsection{Privacy Attacks/Threats}

In ML, privacy attacks and threats refer to techniques or strategies used by adversaries to compromise the privacy of individuals or sensitive data during the training or inference phase of ML models. These attacks aim to exploit vulnerabilities in the ML process to gain unauthorized access to private information or infer sensitive attributes of individuals~\cite{rigaki2020survey}. In the FL area, privacy attacks \hl{refer to} attempts by adversaries to compromise data privacy during the training process. These attacks allow extracting sensitive information from local or aggregated global models to infer, reconstruct, or cause data leakage. \hl{In particular, we categorize such attacks based on the following four main dimensions. First, \textit{method of inference} distinguishes between \emph{passive} attacks (e.g., gradient inversion), which rely on observing shared updates without injecting malicious behavior, and \emph{active} attacks (e.g., canary gradient), which manipulate or perturb updates to increase data leakage. Second, \textit{phase affected} differentiates between leaks that occur predominantly during training (such as gradient-based reconstruction) and those emerging at inference time (like membership inference on final model outputs). Third, \textit{adversary's role} clarifies whether an attacker is an \emph{insider}--a legitimate FL client with access to local computations--or an \emph{outsider} who intercepts or eavesdrops on communication, for instance, through man-in-the-middle tactics. Lastly, the \textit{attack scope} specifies whether an attack is \emph{single-round}, occurring once (e.g., a single instance of eavesdropping), or \emph{multi-round}, gradually accumulating sensitive information over multiple iterations (as in repeated gradient inversion attempts). Thus, Table}~\ref{tab:privacy-categorization}\hl{ summarizes how each known privacy threat fits into these four dimensions. When a threat spans multiple categories (for example, exhibiting both passive and active behaviors), we explicitly mark that overlap.}

\begin{table*}[ht]
\centering
\caption{\hl{Categorization of Privacy Attacks in FL}}
\label{tab:privacy-categorization}
\resizebox{\textwidth}{!}{
\begin{tabular}{@{}lcccccc@{}}
\toprule
\textbf{Privacy Attack} & \textbf{Method (Passive vs.\ Active)} & \textbf{Phase Affected} & \textbf{Adversary's Role} & \textbf{Attack Scope} & \textbf{Goal / Effect} \\ 
\midrule
Gradient Inversion 
    & Passive 
    & Training 
    & Insider 
    & Multi-Round 
    & Reconstruct private data  \\

Gradient Suppression 
    & Active 
    & Training 
    & Insider 
    & Multi-Round 
    & Amplify data leakage patterns \\

Membership Inference 
    & Passive or Active 
    & Inference 
    & Outsider / Insider 
    & Single-Round 
    & Check if a data sample was used in training  \\

Canary Gradient 
    & Active 
    & Training 
    & Insider 
    & Multi-Round 
    & Insert small triggers to deduce sensitive info  \\

Model Inconsistency 
    & Active 
    & Training 
    & Insider (Server) 
    & Single-Round 
    & Compare user updates across different models  \\

GAN-based Inference 
    & Active 
    & Training 
    & Insider 
    & Multi-Round 
    & Generate synthetic data that reveals distribution  \\

Eavesdropping 
    & Passive or Active 
    & Training / Inference 
    & Outsider 
    & Single-Round or Multi-Round 
    & Intercept model updates or network traffic  \\

Unintentional Data Leakage 
    & Passive 
    & Training 
    & Outsider / Insider 
    & Single-Round or Multi-Round 
    & Exploit unintended gradient exposure\\

\bottomrule
\end{tabular}
}
\end{table*}

The following sections explore the most relevant attacks on privacy in FL by defining their nature, objectives, consequences, and examples proposed in the literature.

\subsubsection{Gradient Manipulation}

Gradient manipulation in FL involves exploiting shared gradients to infer or reconstruct sensitive data, posing significant privacy risks. This includes techniques like gradient inversion, reconstruction through inference, and canary attacks, highlighting vulnerabilities in FL's gradient-sharing mechanisms.

\paragraph{Gradient inversion attacks} \hl{These attacks exploit gradients or weight updates shared during the aggregation process in FL to reconstruct private data, posing significant privacy risks}~\cite{neto2023survey}\hl{. These attacks typically leverage optimization techniques or linear relationships between gradients and inputs to infer sensitive information. For instance, Kariyappa et al.}~\cite{kariyappa2023cocktail}\hl{ introduced the Cocktail Party Attack (CPA), which uses independent component analysis to recover private inputs from aggregated gradients, demonstrating its scalability to large batch sizes. This highlights how gradient inversion can compromise privacy even in high-dimensional settings. Li et al.}~\cite{li2022e2egi}\hl{ proposed the End-to-End Gradient Inversion (E2EGI) attack, which iteratively reconstructs training data by reversing gradients, showcasing its potential to breach privacy across multiple iterations. Pasquini et al.}~\cite{pasquini2023security}\hl{ further explored two variants: a passive optimization-based approach that infers private training sets without active interference and an active attack that manipulates model updates to amplify privacy leakage. These methods underline the nuanced mechanisms attackers employ to exploit gradients.

The consequences of gradient inversion attacks are severe. In real-world scenarios, such attacks can expose sensitive medical images or financial records used in FL systems, violating privacy regulations and enabling misuse}~\cite{neto2023survey}\hl{. However, Huang et al.}~\cite{hatamizadeh2023gradient} \hl{argue that practical risks may be mitigated by factors like large batch sizes and local iterations, which reduce reconstruction fidelity. Similarly, Boenisch et al.}~\cite{boenisch2023curious}\hl{ observed that gradient inversion often suffers from local minima and requires extensive iterations for meaningful data recovery, limiting its feasibility in some production environments.} \hl{Implementation typically requires intercepting aggregated gradients and running a local optimization loop that refines random inputs until the gradients match observed signals. Encrypting or clipping gradients partially hinder this by reducing the attacker's visibility or precision, though some accuracy trade-offs may arise.}

\hl{In contrast, \emph{gradient suppression attacks} involve maliciously suppressing gradients during aggregation to manipulate global model updates}~\cite{pasquini2023security}\hl{. By isolating individual updates, attackers can amplify specific patterns in user data, increasing exposure risks. Such attacks can infer the presence of specific data points in user datasets, enabling targeted tracking}~\cite{pasquini2022eluding}\hl{. While their mechanisms differ from gradient inversion, suppression attacks similarly exploit vulnerabilities in gradient-sharing protocols.} \hl{Implementation of gradient suppression often involves intercepting or nullifying certain gradient components before sending them to the server, typically by modifying the local backward pass. A partial defense strategy is to rely on cryptographic checks that ensure gradient consistency across dimensions, thereby preventing an attacker from selectively masking or removing critical features.}

\paragraph{Reconstruction through inference}
\hl{It is a privacy-threatening scenario where an adversary attempts to reconstruct or infer sensitive information about the training data of individual clients by analyzing the model updates or outputs shared during the FL process}~\cite{qammar2022federated}\hl{. Such attacks exploit the inherent vulnerability of gradient-sharing mechanisms in FL. For instance, adversaries can reverse-engineer specific data points or patterns from gradients using techniques like gradient inversion, as demonstrated by Chen et al.}~\cite{chen2023feddef}\hl{. They identified two distinct types of reconstruction attacks. The first, called extraction attack, focuses on accurately reconstructing a single training sample with minimal computational cost. This attack leverages advanced optimization techniques to improve reconstruction accuracy, posing significant risks to data privacy. The second type, manipulating reconstructed data, allows adversaries to recover private training data and labels from gradients and subsequently modify this data to execute targeted attacks on models. For example, in healthcare FL applications, attackers could reconstruct sensitive medical images shared across hospitals and manipulate them to mislead diagnostic models}~\cite{dahlgaard2022analysing}. \hl{Attackers mostly rely on final model outputs or partial gradient snapshots for offline reconstruction, requiring minimal changes to the FL pipeline. Defensive measures like gradient masking or cryptographic aggregation reduce the granularity of the information available, limiting reconstruction success.}

\paragraph{Canary Gradient}
\hl{A canary gradient attack is a privacy breach in FL where an attacker exploits gradients or weight updates shared during the aggregation process to infer sensitive information. Its name originates from using canaries in coal mines to detect poisonous gases}~\cite{pasquini2022eluding}\hl{. In this attack, the adversary injects small, carefully crafted perturbations into a client's gradients or weight updates and observes the server's response to deduce private data. For instance, such attacks have been shown to reconstruct sensitive client data under certain conditions, raising concerns about FL's privacy guarantees}~\cite{maddock2022canife}\hl{. Maddock et al.}~\cite{maddock2022canife}\hl{ propose CANIFE, a method to evaluate empirical privacy risks in FL by introducing adversarially crafted "canary" samples. These samples are used to measure model exposure to privacy breaches, revealing that the empirical per-round privacy loss is significantly tighter than theoretical bounds. This approach highlights vulnerabilities in FL systems, such as susceptibility to gradient inversion attacks in real-world scenarios, which theoretical DP guarantees may underestimate. By offering a realistic assessment of privacy risks, CANIFE underscores gaps in current defenses and emphasizes the importance of robust threat models for FL.} \hl{To carry out a canary attack, an adversary could inject imperceptible 'signatures' into local gradients, then checks if these signatures reappear in the global model's updates. Clipping (see Section}~\ref{subsec:privacy_defenses}\hl{ or encrypting gradients dilutes such signatures, minimizing the attacker's ability to confirm the presence of sensitive data.}

\subsubsection{Membership Inference}
\hl{In this privacy threat, an adversary seeks to determine whether a specific data point was part of a client's training dataset used in the FL process. Such an attack primarily aims to verify the membership status of individual data points, discerning whether they belong to a client's private training data}~\cite{yin2021comprehensive}\hl{. This breach of privacy may lead to the disclosure of identities or sensitive attributes, undermining the confidentiality and anonymity of data contributors. Zhang et al.}~\cite{zhang2023agrevader}\hl{ highlights two types of membership inference attacks with distinct mechanisms and implications. Poisoning membership inference attacks involve adversaries injecting carefully crafted malicious samples into the training data to detect membership. For instance, by observing how poisoned examples alter model loss, attackers can infer membership, posing severe risks in healthcare FL systems where patient data is highly sensitive. Black-box membership inference attacks, such as Memguard}~\cite{jia2019memguard}\hl{, operate without direct access to training data or models. These attacks generate adversarial queries to exploit model predictions and infer membership, which could compromise user anonymity in recommendation systems. Pasquini et al.}~\cite{pasquini2023security}\hl{ emphasize that adversaries can leverage model updates or query responses to enhance their guesses. The harm caused by these attacks extends beyond privacy breaches, as they can facilitate further privacy threats like attribute inference attacks, creating a cascading effect on the overall security of FL systems.} \hl{Specifically, membership inference often queries the global model's confidence scores for specific inputs. Small modifications to the local or server-side scripts can track these score patterns, exposing training-set membership. Defensive techniques such as local DP or randomizing confidence outputs inhibit the attack's reliability.}

\subsubsection{Model Inconsistency}
\hl{This attack exploits a vulnerability in the FL protocol caused by incorrect usage of secure aggregation and a lack of parameter validation. Specifically, a malicious server distributes different versions of the model to different users within the same training round. The server can analyze behavioral differences in user updates, even though these updates are securely aggregated. The attack leverages the fact that varying model parameters can induce detectable differences in gradient updates, which may reveal sensitive training data. For instance, Pasquini et al.}~\cite{pasquini2022eluding}\hl{ demonstrated that this approach enables inference of private information regardless of the number clients. Real-world implications include risks to applications like healthcare and IoT, where sensitive data is prevalent. Zhang et al.}~\cite{zhang2023personalized}\hl{ further noted that inconsistencies between global and local models could reflect attack-related information, potentially guiding personalized FL algorithms to improve fault diagnosis accuracy. These findings underscore the severe privacy risks of model inconsistency attacks and their potential to compromise FL systems at scale.} \hl{The implementation of such an attack only requires slight server-side changes--assigning slightly different model parameters to each client in a single round. A recommended mitigation is verifying model consistency across clients or leveraging secure multi-party aggregation to ensure identical parameter distributions.}

\subsubsection{GANs-based Inference}
\hl{A GANs-based inference attack in FL uses GANs to infer sensitive information about the training data held by individual clients. This attack aims to create a generator network that can produce data samples indistinguishable from the data used for training in the local client models}~\cite{neto2023survey}\hl{. The attacker can effectively determine whether a specific data point was part of a client's training dataset, conducting membership inference. For example, in medical diagnosis scenarios, such attacks could allow a malicious client to infer sensitive patient conditions from gradients shared during FL updates}~\cite{huang2023mitigating}\hl{. The consequences of such attacks are severe, as they breach privacy by revealing which data points were used during training, leading to potential misuse or discrimination risks. Huang and Xiang}~\cite{huang2023mitigating}\hl{ introduce Cross-Client GANs (C-GANs) attacks, where a malicious client reconstructs samples resembling the distribution of other clients' training data. This enables adversaries to compromise privacy by leaking benign clients' sensitive data, as demonstrated in experiments involving reconstructed images from medical datasets.} \hl{Attackers usually train a local generator alongside the global model, refining synthetic samples that mimic real client data. By limiting gradient visibility with cryptographic or noisy protocols, such as DP, defenders hamper the generator's ability to converge on sensitive distributions.}

\subsubsection{Eavesdropping}
\hl{In this threat, an attacker intercepts the communication between the clients and the central server. It is done by sniffing the network traffic or by compromising the devices of the clients or the server}~\cite{xu2021else}\hl{. The objectives of an eavesdropping attack often include stealing the FL model, inferring sensitive information from the client's data, and disrupting the FL process. Guo et al.}~\cite{guo2023federated}\hl{ identifies four types of eavesdropping attacks: Passive Eavesdropping, where an attacker monitors communication without altering data; Active Eavesdropping, involving data modification, such as injecting malicious gradients; Man-in-the-Middle (MitM) Attacks, which intercept and relay messages, posing significant risks to data integrity and confidentiality; and Network Sniffing, capturing network traffic to extract sensitive information.} \hl{Practical eavesdropping often exploits unsecured Wi-Fi or inadequate encryption. An attacker needs little more than packet-capture tools to observe local updates. Configuring Transport Layer Security (TLS) or implementing fully HE (FHE) for parameter exchanges effectively thwarts passive intercepts.}

\subsubsection{Unintentional Data Leakage}
The latter is not precisely an attack, but it is more of a vulnerability that attackers can exploit once discovered. It occurs when private training data leaks through the gradient-sharing mechanism deployed in FL systems. The objective of this is to recover batch data from the shared aggregated gradients. The latter can be catastrophic and lead to users' private data reconstruction by eavesdropping on shared gradients. The risk of confidential data leaking from the training gradients in standard FL, especially the vertical case, is high. For Nair et al.\cite{nair2023privacy}, these vulnerabilities are a concern in FL due to the potential for data leakage and adversarial attacks during gradient transfer operations. These threats can occur when gradients are transferred between clients in the FL system. Ziz et al.~\cite{jin2021cafe} introduce the CAFE (catastrophic data leakage in vertical FL) attack, an advanced data leakage attack in FL that aims to recover private data from shared aggregated gradients. It addresses the limitations of existing approaches regarding scalability and theoretical justification for data recovery. The attack algorithm consists of three steps: (1) Recovering the loss gradients concerning the outputs of the first fully connected (FC) layer. (2) Using the recovered gradients as a learned regularizer to improve the performance of the data leakage attack. (3) Using the updated model parameters to perform the data leakage attack. \hl{Such leakage arises when partial gradient details or intermediate layer outputs inadvertently reveal private features. Minimizing or masking these signals (via secure aggregation or randomization) lowers the precision with which attackers can reassemble original training data.}

\begin{figure}[ht]
  \centering
  \includegraphics[width=\linewidth]{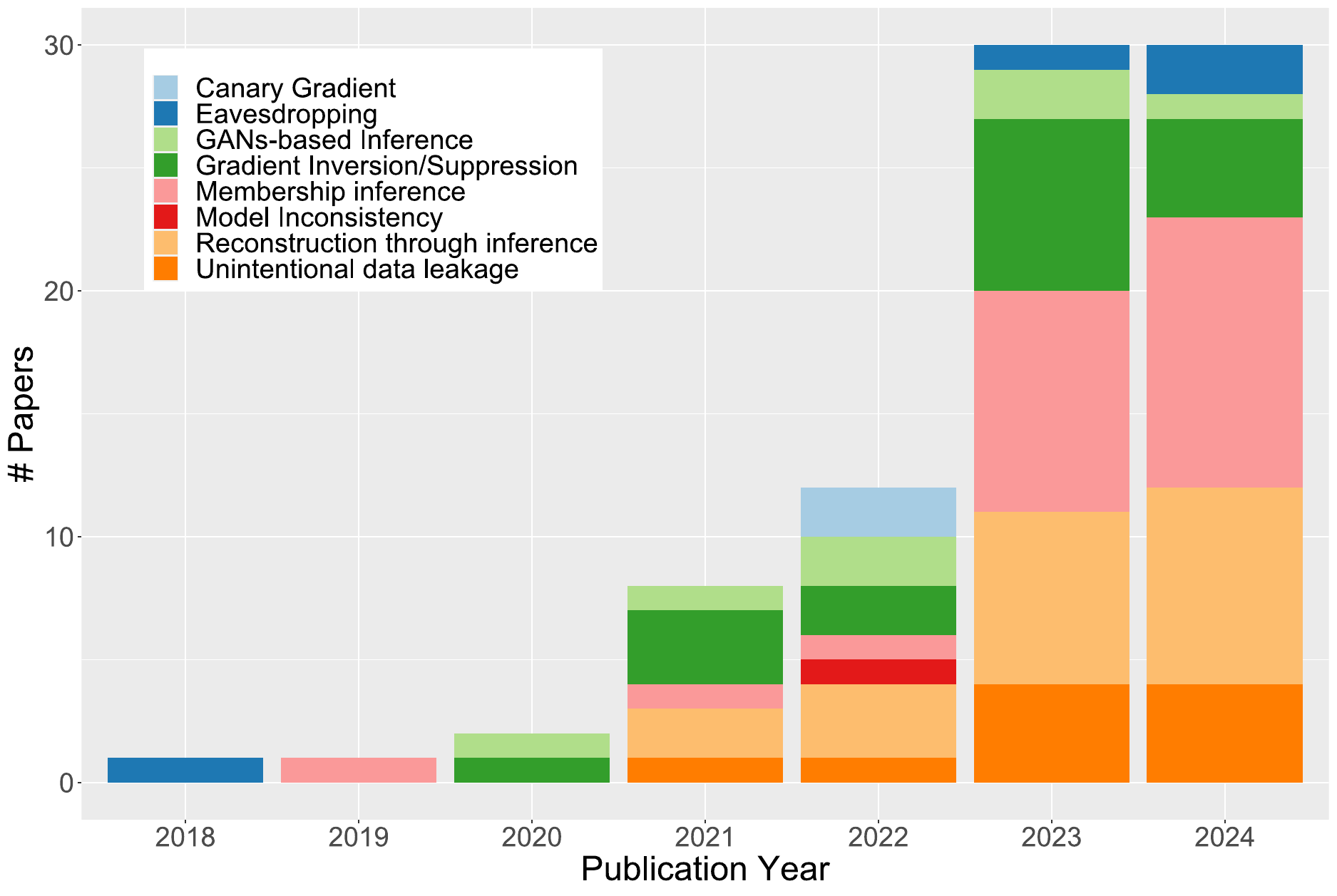}
  \caption{Papers related to privacy attacks over time}
  \label{fig:privacy_attacks_in_time}
\end{figure}

Leveraged on the literature review done, we retrieved a list of privacy threats included in the current literature. Fig.~\ref{fig:privacy_attacks_in_time} demonstrates a notable increase in privacy attack research publications, reflecting heightened awareness and concern in this field. The early years of 2018 until 2020 saw minimal research output with a narrow focus on attack types. However, a marked shift occurred since 2021, with a substantial rise in the quantity and range of studies. Attacks like Membership inference and Reconstruction through inference gained significant traction, particularly in 2023 and 2024. Concurrently, research on GANs-based Inference and Gradient Inversion/Suppression expanded considerably, with a noticeable peak in 2024. The emergence of Unintentional data leakage as a research focus in recent years adds to the diversifying landscape. The year 2024 stands out with the most comprehensive coverage across various attack categories, indicating an intensified focus on privacy protection research. This progression highlights the dynamic nature of privacy threats and showcases the academic community's proactive stance in addressing evolving challenges in data privacy safeguarding.

\subsection{Privacy Defenses}\label{subsec:privacy_defenses}
Simultaneously, as the range and intricacy of attacks and threats on FL grow, novel defenses are also emerging to counteract their harmful impacts~\cite{rodriguez2023survey}. The following sections explore the most relevant defenses on privacy in FL by defining their concept, advantages, disadvantages, and the attacks each type of defense can defend against, proposed in the most recent and pertinent literature.

\subsubsection{Zero-knowledge Proof}
Zero-knowledge proof-based FL (ZKP-FL) scheme leverages zero-knowledge proof for both the computation of local data and the aggregation of local model parameters, aiming to verify the computation process without requiring the plaintext of the local data. Xing et al.~\cite{xing2023zero} provided that on a blockchain, ZKP-FL allows clients in an FL system to prove to the central server that they have computed the correct model updates without revealing their underlying data. It helps to protect against attacks that aim to infer sensitive information from the federated model or the FL process. The FLAG framework~\cite{bangalore2023flag} utilizes ZKP-FL techniques to provide secure computation without revealing sensitive information. It protects against data leakage, inference attacks, and unauthorized access to sensitive information. It provides a lightweight and efficient framework for secure aggregation in FL. However, implementing ZKP-FL protocols requires additional computational resources, which may introduce complexity and overhead due to the need for safe communication and encryption. 

\subsubsection{Oblivious Transfer}
Oblivious Transfer (OT) in FL refers to a cryptographic protocol that allows a client to obtain one out of multiple potential values from a server without revealing the chosen value to the server. OT ensures privacy and confidentiality in FL by enabling clients to securely access and retrieve information from the server without compromising sensitive data. Rathee et al.~\cite{rathee2023elsa} introduced ELSA (Ensemble Learning with Semi-honest Aggregators) as a defense mechanism in FL to protect the privacy of individual gradients during aggregation. ELSA employs l2-norm bounding to defend against boosted gradients from malicious clients. It consists of multiple layers that incorporate semi-honest and adversarial privacy defenses. ELSA's l2-norm protection is relatively simple compared to other securities, making it suitable for working over secret shares. ELSA cannot guarantee fairness, as it cannot distinguish between a malicious server and a malicious client, potentially leading to some honest clients' inputs not being used in the computation. ELSA's privacy defenses aim to protect the privacy of individual gradients during aggregation, limiting information leakage from the global aggregate.

\subsubsection{Homomorphic Encryption (HE)}

HE is a cryptographic technique that enables computations to be performed on encrypted data, producing an encrypted result that, when decrypted, matches the result of the same operations performed on the (original) plaintext. Singh et al.~\cite{singh2022framework} discuss using HE to protect crucial data in the healthcare system, allowing computations to use encrypted data without decrypting it, preserving privacy. It enables secure data sharing and analysis in FL without revealing sensitive information to the central server or other clients. Nevertheless, it is a complex technology that requires specialized knowledge and infrastructure for implementation. The solution protects against unauthorized access, data leakage, and inference attacks. HE is a critical component of the SoK~\cite{mansouri2023sok} defense strategy in FL. It offers robust privacy guarantees by ensuring data remains encrypted throughout the computation process. However, it can introduce notable drawbacks, including computational overhead and increased communication costs due to the complexity of processing encrypted data. It serves as a robust defense against various threats, including eavesdropping and data inference attacks, ultimately enhancing the privacy of FL systems.

\subsubsection{Secret Sharing}
Secret sharing (SS) is used in FL to distribute sensitive information, such as model parameters, among multiple clients. It involves dividing the secret into shares and distributing them to different clients, ensuring that no single client can access the complete secret. tMK-CKKS~\cite{du2023efficient} with secret sharing provides information-theoretic security, making it impervious to collusion attacks by up to \emph{t-1} clients working in concert with the server. Even when many clients join forces, they cannot deduce details about the master's secret. It involves the distribution of the master public key among all clients for encryption purposes. Furthermore, individual secret keys for each client are generated using a linear secret-sharing scheme. Notably, the decryption of aggregated ciphertexts necessitates the cooperation of only a specific threshold value, \emph{t} clients. This careful balance of secret sharing and threshold requirements enhances the FL system's privacy.

\subsubsection{MPC}
MPC allows multiple clients to collaboratively compute a function on their private inputs without disclosing those inputs to one another. Bangalore et al.~\cite{bangalore2023flag} proposed FLAG that scales to 1000s of clients, requires only a constant number of rounds, outperforms prior work in computational cost, and has competitive communication cost. However, it may introduce computational overhead due to the need for secure protocols and encryption. It helps defend against attacks that aim to compromise the privacy of user-held data during the aggregation process in FL. Mansouri et al.~\cite{mansouri2023sok} proposed SoK, a defense mechanism in FL that focuses on the MPC of data from multiple sources without revealing individual inputs. It provides privacy protection and prevents adversaries from inferring sensitive information. SOK defense offers advantages in FL, like ensuring that individual data contributions remain confidential, making it difficult for attackers to carry out these attacks by obfuscating individual data contributions. Some downsides may include additional computational overhead and communication costs, specialized cryptographic knowledge, and careful design to ensure efficiency and scalability. SoK was specially designed to prevent membership inference and data reconstruction attacks.

\subsubsection{DP}
Nagy et al.~\cite{nagy2023privacy} proposed a privacy-preserving FL framework for natural language processing incorporating local differential privacy (LDP) as a robust defense mechanism. LDP safeguards the privacy of individual data contributions, introducing noise to the model updates before sharing it with the server. The critical advantage of LDP is that it is exceedingly challenging for potential attackers to infer sensitive information about individual data contributors. However, introducing noise can impact the accuracy of the trained ML models, necessitating a careful balance between privacy preservation and model utility. LDP is a defense against attacks to uncover sensitive information about individual data contributors. Dynamic differential privacy (DDP), proposed by Guo et al.~\cite{guo2023federated}, involves dynamically adjusting the privacy budget and noise scale during model training, allowing for higher-quality models with a fixed privacy budget. It helps to get higher quality models and real-time privacy tracking, preventing the privacy budget from being exceeded, which could lead to the leakage of sensitive information. One drawback of this method is that it requires careful adjustment and injection of noise in each iteration, which adds complexity to the FL process. It is particularly effective at defending against eavesdropping attacks.

\subsubsection{Gradient Clipping}
Gradient-based defenses in FL involve modifying the gradients during training to protect against adversarial attacks. These defenses aim to make the model more robust by perturbing the gradients or adding noise. Chen et al.~\cite{chen2023feddef} proposed FedDef as an optimization-based input perturbation defense in FL that aims to preserve privacy and FL model performance by transforming private data into pseudo data that is dissimilar to the original data while maintaining similar gradients. Users download the global model from the server and use FedDef during local training to transform their data and gradients. However, the computational and memory overhead of FedDef needs to be considered, compared with HE in terms of performance. FedDef defends against reconstruction attacks, such as inversion attacks, extraction attacks, and attacks that manipulate reconstructed data. \hl{Gradient clipping restricts the size of updates from individual clients, ensuring that large updates from skewed or non-IID data do not disproportionately impact the model performance.}

Li et al.~\cite{li2022auditing} study \emph{gradient clipping} and \emph{sparsification} as defense mechanisms in FL. \emph{Gradient clipping} involves setting a threshold for the magnitude of gradients during training and scaling them down if they exceed the threshold. This technique helps mitigate privacy leakage risks by controlling gradient magnitudes, making it harder to infer sensitive information from them. However, striking the right balance between privacy and model performance is crucial, as overly stringent thresholds can hinder learning. On the other hand, \emph{gradient sparsification} enhances privacy by selectively transmitting only a subset of gradients that exceed a specified threshold. The latter reduces the information shared with the central server, minimizing the risk of privacy breaches. While it offers strong privacy protection, it may introduce computational overhead and affect the learning process by discarding some information.

\subsubsection{Personalized Solutions} The references reviewed offer tailor-made solutions created to improve the security of FL environments. The most relevant are mentioned below.

\paragraph{CrowdFL}
CrowdFL~\cite{zhao2022crowdfl,zhao2021crowdsensing} is an innovative approach that combines mobile crowdsensing (MCS) with FL to address privacy concerns while harnessing the computational power of clients. In this system, participants can perform local data processing using the FL framework, ensuring that sensitive sensing data remains on their clients. Only encrypted training models are uploaded to the server, preserving clients' privacy. CrowdFL offers scalability by leveraging MCS's large-scale data collection capabilities and reduces deployment costs by eliminating the need for extensive centralized infrastructure. This integration of FL into MCS enhances privacy and makes it a cost-effective and scalable solution for privacy-preserving mobile crowdsensing applications.

\paragraph{Soteria}
Soteria~\cite{sun2021soteria,chen2023feddef} is a defense mechanism proposed against model inversion attacks in FL. The defense focuses on perturbing data representation to severely degrade the quality of reconstructed data while maintaining FL performance. It aims to improve the privacy of FL systems by addressing data representation leakage from gradients, which has been identified as the essential cause of privacy leakage. After applying the protection, it provides a certified robustness guarantee to FL and a convergence guarantee to FedAvg. The privacy of the FL system is significantly improved with the implementation of Soteria defense. Soteria is designed to defend against model inversion attacks in FL, specifically the deep leakage from gradients (DLG) and gradient stalking (GS) attacks. These attacks aim to infer private data by exploiting the vulnerability of FL to inference attacks. \hl{Soteria ensures that the perturbed representations remain similar to the true representations for effective learning but degrade the quality of reconstructed data, mitigating the exacerbated privacy risks caused by non-IID data distributions.}

\paragraph{FLTrust}~\cite{cao2020fltrust,zhang2023agrevader}: FLTrust is a defense mechanism designed to enhance the privacy of FL, aiming to detect and mitigate malicious clients in the FL process by evaluating their trustworthiness based on their behavior and contributions to the model training process. FLTrust utilizes trust scores to assess the reliability of clients and make informed decisions regarding their inclusion in the FL system. It relies on trust scores, which may not always accurately reflect the true intentions of clients. False positives or false negatives in trust evaluation can impact the fairness and effectiveness of the defense mechanism. FLTrust defends against attacks involving malicious clients in FL, such as Byzantine poisoning adversarial attacks, local models, and poisoning attacks in Byzantine-robust FL. This method addresses non-IID data issues (see Section~\ref{sec:non-iid-impact}) using a trusted server-side dataset to evaluate and assign trust scores to client updates, ensuring that malicious or biased updates are down-weighted.


\paragraph{RoFL}~\cite{burkhalter2021rofl,lycklama2023rofl}: Robustness of secure FL (RoFL) is an FL system that incorporates constraints on clients' updates to mitigate severe attacks. It extends secure aggregation with privacy-preserving input validation. RoFL efficiently includes conditions such as norm bounds on clients' updates and provides secure FL protocols in the single-server setting. RoFL can enforce restrictions such as $L_2$ and $L_\infty$ bounds on high-dimensional encrypted model updates. RoFL achieves practicality even at a large scale. However, it incurs considerable overhead in terms of computational resources and, notably, bandwidth. \hl{It also uses a fairness-aware optimization approach to ensure balanced contributions from clients, improving overall model performance on non-IID datasets.}

\paragraph{Prio}~\cite{corrigan2017prio,rathee2023elsa}: It primarily focuses on preserving privacy while collecting essential aggregate statistics. This system leverages specialized zero-knowledge proofs, known as SNIPs, to enforce diverse defenses against malformed gradients while ensuring the confidentiality of clients' data against the influence of at most one malicious server. It not only safeguards sensitive information but also provides the flexibility to implement various defenses against irregular gradients, enhancing the overall security of the FL process. However, it's essential to consider potential drawbacks, such as the potential for computational overhead and increased communication costs due to using zero-knowledge proofs. In essence, Prio defense plays a crucial role in defending FL against attacks that manipulate gradients and compromise the integrity of aggregated statistics, all while preserving privacy.

\begin{table}[htbp]
    \centering
    \resizebox{\columnwidth}{!}{
    \begin{tabular}{@{}lccccccccc@{}}
      & \multicolumn{8}{c}{\textbf{Attacks/Threats}} \\
        \cmidrule{2-9}
        \textbf{Defenses} & \rotatebox{90}{\textbf{\makecell{Gradient  \\ Inversion/suppression}}} 
        & \rotatebox{90}{\textbf{\makecell{Reconstruction \\ through inference}}} 
        & \rotatebox{90}{\textbf{Membership inference}} 
        & \rotatebox{90}{\textbf{Canary Gradient}} 
        & \rotatebox{90}{\textbf{Model Inconsistency}} 
        & \rotatebox{90}{\textbf{GANs-based inference}} 
        & \rotatebox{90}{\textbf{Eavesdropping}} 
        & \rotatebox{90}{\textbf{\makecell{Unintentional \\data leakage}}} \\
        \midrule
        \textbf{Zero-knowledge proof}            & & & & & \textcolor{teal}{\ding{52}} & & & \\
        \textbf{Oblivious Transfer}              & & \textcolor{teal}{\ding{52}} & & & & & & \\
        \textbf{Homomorphic encryption}          & \textcolor{teal}{\ding{52}} & & & & & & & \textcolor{teal}{\ding{52}} \\
        \textbf{Secret sharing}                  & \textcolor{teal}{\ding{52}} & & & & & & & \\
        \textbf{MPC}   & \textcolor{teal}{\ding{52}} & \textcolor{teal}{\ding{52}} & \textcolor{teal}{\ding{52}} & & & & & \\
        \textbf{Differential privacy}            & \textcolor{teal}{\ding{52}} & \textcolor{teal}{\ding{52}} & \textcolor{teal}{\ding{52}} & \textcolor{teal}{\ding{52}} & & & & \textcolor{teal}{\ding{52}} \\
        \textbf{Gradient clipping}                  & & & & & \textcolor{teal}{\ding{52}} & & & \\
        \textbf{Additive Noise}                  & \textcolor{teal}{\ding{52}} & & \textcolor{teal}{\ding{52}} & \textcolor{teal}{\ding{52}} & & & & \\
        \textbf{Personalized solutions}          & \textcolor{teal}{\ding{52}} & & \textcolor{teal}{\ding{52}} & & \textcolor{teal}{\ding{52}} & & & \\
        \bottomrule
    \end{tabular}
    }
    \caption{Relationship of defense mechanisms and attacks
for private FL}
    \label{tab:attacks-defense-privacy-comparison}
\end{table}

\hl{Table}~\ref{tab:attacks-defense-privacy-comparison}\hl{ highlights the intricate interplay between privacy threats and defense mechanisms in FL, revealing critical gaps and strengths in existing approaches. Notably, DP and MPC emerge as versatile solutions, countering a broad spectrum of attacks, including gradient inversion, membership inference, and unintentional data leakage. However, their effectiveness often comes at the cost of computational overhead and accuracy trade-offs, limiting real-world adoption. HE and secret sharing offer robust protection against gradient-based attacks but are less effective against inference threats like model inconsistency or GAN-based inference. Personalized solutions provide targeted defenses but lack the generalizability required for large-scale deployments. Moreover, attacks such as model inconsistency and GAN-based inference remain under-defended, signaling a need for novel countermeasures. These observations suggest that while FL privacy defenses are evolving, a holistic approach that balances security, efficiency, and adaptability remains a key challenge.}

\emph{Lessons learned:} The analysis of privacy threats and defenses in FL highlights a dynamic and evolving landscape. Attacks such as gradient inversion, membership inference, and reconstruction-based techniques continue to pose significant risks, \textcolor{black}{exposing systemic weaknesses in current gradient-sharing protocols and the lack of standardized privacy auditing frameworks.}. The rise of sophisticated attack strategies, including adversarial GANs and model inconsistency exploits, emphasizes the need for stronger, adaptive defenses \textcolor{black}{, particularly those capable of operating under heterogeneous client behaviors and long training cycles}. On the defense side, privacy-preserving techniques like DP, HE, and MPC show promise but often introduce computational overhead and may degrade model utility. \textcolor{black}{Despite these limitations, hybrid protocols (e.g., DP-MPC) offer promising trade-offs, yet their scalability and deployment in real-world FL settings remain largely underexplored. Another challenge is the fragmented evaluation of defense strategies; indeed, most studies lack unified benchmarks or attack coverage, making it difficult to assess robustness across multiple threat vectors.} Personalized solutions, such as FLTrust and Soteria, offer targeted protection against specific attack vectors but may require careful tuning to balance security and efficiency. \textcolor{black}{Lastly, while many efforts focus on protecting server-side aggregation or model parameters, client-side privacy preservation (e.g., during local training or device compromise) remains insufficiently addressed, creating opportunities for attack vectors beyond the current scope of most defenses.}

\subsection{Comparative Analysis of Defenses for Private FL}

\hl{In this subsection, we examine quantitative and experimental research to assess how well previous privacy defenses perform under particular attack scenarios}~\cite{zahri2023empirical, zhu2023secure, das2025communication, zhang2023advancing}\hl{. DP is effective against inference attacks by adding noise to gradients, but it often reduces accuracy, especially for underrepresented classes; newer methods like DP-MPC improve efficiency significantly (16-182x faster) while maintaining privacy guarantees. HE ensures strong confidentiality by enabling computations on encrypted data, achieving high accuracy (e.g., ~99.95\% on MNIST) but at a significant computational cost. MPC, particularly when combined with DP, enhances communication efficiency (56-794x) and speed, making it suitable for both privacy and efficiency scenarios. Gradient Clipping stabilizes training and reduces the risks of exploding gradients, but can slightly degrade performance if thresholds are too restrictive. ZKPs provide strong privacy and verifiability in trustless environments, such as blockchain-based FL, but can be computationally expensive. DP-MPC is recommended for applications prioritizing privacy without excessive overhead due to its balance of efficiency and privacy. Despite its cost, HE is ideal for accuracy-critical tasks like medical imaging, while gradient clipping with DP is effective for resource-constrained scenarios. In decentralized systems requiring trustless operations, ZKPs are valuable but should be used selectively due to their computational demands.
}

\hl{FL inherently involves trade-offs between privacy, security, and model performance. Strengthening privacy mechanisms often reduces model utility, while enhancing security may increase computational overhead}~\cite{nasr2018comprehensive, tramer2020differentially}\hl{. Therefore, selecting appropriate techniques depends on the specific application needs—whether prioritizing efficiency, accuracy, or privacy. Understanding these trade-offs is essential for designing robust and practical FL systems}~\cite{mohammadi2024balancing}.

\section{FL Frameworks}
\label{sec:fl_frameworks}
FL has witnessed the emergence of several frameworks designed to facilitate its application and address various aspects of privacy and security. \hl{Table}~\ref{tab:frameworks_compare} \hl{compares multiple features evaluated for the mentioned FL frameworks. The comparison involves the privacy and security support, attack simulation capacity, FL implemented types, and documentation and tutorials provided for the users. This comparative analysis enables us to provide insights into the strengths and weaknesses of each FL framework in terms of privacy, security, functionality, and user-friendliness, aiding researchers and practitioners in selecting the most suitable framework for their specific needs. The following paragraphs overview the most relevant FL frameworks, highlighting their characteristics, advantages, and limitations. At the end of the section, we also provide some lessons learned from analyzing the frameworks.}

\begin{table*}[t!]
  \caption{FL frameworks features comparison (\textcolor{teal}{\ding{52}}: Complete, \textcolor{orange}{\ding{117}}: Under development/incomplete, \textcolor{purple}{\ding{56}}: Unknown/Not implemented)}
  \label{tab:frameworks_compare}
  \resizebox{\textwidth}{!}{\begin{tabular}{lccccccccccccc}
    \toprule
     & \textbf{CRYPTEN} & \textbf{FATE} & \textbf{FEDML} & \textbf{FEDSCALE} & \textbf{FL AND DP} & \textbf{FLOWER} & \textbf{FLUTE} & \textbf{NVFLARE} & \textbf{OPENFL} & \textbf{PaddleFL} & \textbf{PYSYFT} & \textbf{TFF} & \textbf{XFL} \\
    \midrule
    \textbf{Privacy and security support} & \textcolor{orange}{\ding{117}} & \textcolor{orange}{\ding{117}} & \textcolor{teal}{\ding{52}} & \textcolor{orange}{\ding{117}} & \textcolor{teal}{\ding{52}} & \textcolor{teal}{\ding{52}} & \textcolor{orange}{\ding{117}} & \textcolor{orange}{\ding{117}} & \textcolor{orange}{\ding{117}} & \textcolor{orange}{\ding{117}} & \textcolor{teal}{\ding{52}} & \textcolor{orange}{\ding{117}} & \textcolor{orange}{\ding{117}} \\
    \midrule
    \hspace{2mm}\ding{101} Differential privacy & \textcolor{purple}{\ding{56}} & \textcolor{purple}{\ding{56}} & \textcolor{teal}{\ding{52}} & \textcolor{teal}{\ding{52}} & \textcolor{teal}{\ding{52}} & \textcolor{teal}{\ding{52}} & \textcolor{teal}{\ding{52}} & \textcolor{teal}{\ding{52}} & \textcolor{teal}{\ding{52}} & \textcolor{teal}{\ding{52}} & \textcolor{teal}{\ding{52}} & \textcolor{teal}{\ding{52}} & \textcolor{teal}{\ding{52}} \\
    \hspace{2mm}\ding{101} FoolsGold & \textcolor{purple}{\ding{56}} & \textcolor{purple}{\ding{56}} & \textcolor{teal}{\ding{52}} & \textcolor{purple}{\ding{56}} & \textcolor{purple}{\ding{56}} & \textcolor{purple}{\ding{56}} & \textcolor{purple}{\ding{56}} & \textcolor{purple}{\ding{56}} & \textcolor{purple}{\ding{56}} & \textcolor{purple}{\ding{56}} & \textcolor{purple}{\ding{56}} & \textcolor{purple}{\ding{56}} & \textcolor{purple}{\ding{56}} \\
    \hspace{2mm}\ding{101} GeoMed & \textcolor{purple}{\ding{56}} & \textcolor{purple}{\ding{56}} & \textcolor{purple}{\ding{56}} & \textcolor{purple}{\ding{56}} & \textcolor{purple}{\ding{56}} & \textcolor{purple}{\ding{56}} & \textcolor{purple}{\ding{56}} & \textcolor{purple}{\ding{56}} & \textcolor{purple}{\ding{56}} & \textcolor{purple}{\ding{56}} & \textcolor{purple}{\ding{56}} & \textcolor{purple}{\ding{56}} & \textcolor{purple}{\ding{56}} \\
    \hspace{2mm}\ding{101} Homomorphic encryption & \textcolor{purple}{\ding{56}} & \textcolor{teal}{\ding{52}} & \textcolor{teal}{\ding{52}} & \textcolor{purple}{\ding{56}} & \textcolor{teal}{\ding{52}} & \textcolor{teal}{\ding{52}} & \textcolor{purple}{\ding{56}} & \textcolor{teal}{\ding{52}} & \textcolor{purple}{\ding{56}} & \textcolor{purple}{\ding{56}} & \textcolor{teal}{\ding{52}} & \textcolor{orange}{\ding{117}} & \textcolor{teal}{\ding{52}} \\
    \hspace{2mm}\ding{101} Krum & \textcolor{purple}{\ding{56}} & \textcolor{purple}{\ding{56}} & \textcolor{teal}{\ding{52}} & \textcolor{purple}{\ding{56}} & \textcolor{purple}{\ding{56}} & \textcolor{teal}{\ding{52}} & \textcolor{purple}{\ding{56}} & \textcolor{purple}{\ding{56}} & \textcolor{purple}{\ding{56}} & \textcolor{purple}{\ding{56}} & \textcolor{purple}{\ding{56}} & \textcolor{purple}{\ding{56}} & \textcolor{purple}{\ding{56}} \\
    \hspace{2mm}\ding{101} Multi-Krum & \textcolor{purple}{\ding{56}} & \textcolor{purple}{\ding{56}} & \textcolor{teal}{\ding{52}} & \textcolor{purple}{\ding{56}} & \textcolor{purple}{\ding{56}} & \textcolor{teal}{\ding{52}} & \textcolor{purple}{\ding{56}} & \textcolor{purple}{\ding{56}} & \textcolor{purple}{\ding{56}} & \textcolor{purple}{\ding{56}} & \textcolor{purple}{\ding{56}} & \textcolor{purple}{\ding{56}} & \textcolor{purple}{\ding{56}} \\
    \hspace{2mm}\ding{101} Norm difference clipping & \textcolor{purple}{\ding{56}} & \textcolor{purple}{\ding{56}} & \textcolor{teal}{\ding{52}} & \textcolor{purple}{\ding{56}} & \textcolor{purple}{\ding{56}} & \textcolor{teal}{\ding{52}} & \textcolor{purple}{\ding{56}} & \textcolor{purple}{\ding{56}} & \textcolor{purple}{\ding{56}} & \textcolor{purple}{\ding{56}} & \textcolor{purple}{\ding{56}} & \textcolor{teal}{\ding{52}} & \textcolor{purple}{\ding{56}} \\
    \hspace{2mm}\ding{101} RFA (geometric median) & \textcolor{purple}{\ding{56}} & \textcolor{purple}{\ding{56}} & \textcolor{teal}{\ding{52}} & \textcolor{purple}{\ding{56}} & \textcolor{purple}{\ding{56}} & \textcolor{purple}{\ding{56}} & \textcolor{purple}{\ding{56}} & \textcolor{purple}{\ding{56}} & \textcolor{purple}{\ding{56}} & \textcolor{purple}{\ding{56}} & \textcolor{purple}{\ding{56}} & \textcolor{purple}{\ding{56}} & \textcolor{purple}{\ding{56}} \\
    \hspace{2mm}\ding{101} Secret Sharing & \textcolor{teal}{\ding{52}} & \textcolor{purple}{\ding{56}} & \textcolor{purple}{\ding{56}} & \textcolor{purple}{\ding{56}} & \textcolor{teal}{\ding{52}} & \textcolor{purple}{\ding{56}} & \textcolor{purple}{\ding{56}} & \textcolor{purple}{\ding{56}} & \textcolor{purple}{\ding{56}} & \textcolor{purple}{\ding{56}} & \textcolor{teal}{\ding{52}} & \textcolor{purple}{\ding{56}} & \textcolor{purple}{\ding{56}} \\
    \hspace{2mm}\ding{101} Secure Aggregation & \textcolor{purple}{\ding{56}} & \textcolor{purple}{\ding{56}} & \textcolor{purple}{\ding{56}} & \textcolor{purple}{\ding{56}} & \textcolor{purple}{\ding{56}} & \textcolor{teal}{\ding{52}} & \textcolor{purple}{\ding{56}} & \textcolor{purple}{\ding{56}} & \textcolor{purple}{\ding{56}} & \textcolor{teal}{\ding{52}} & \textcolor{orange}{\ding{117}} & \textcolor{purple}{\ding{56}} & \textcolor{purple}{\ding{56}} \\
    \hspace{2mm}\ding{101} MPC & \textcolor{teal}{\ding{52}} & \textcolor{teal}{\ding{52}} & \textcolor{purple}{\ding{56}} & \textcolor{purple}{\ding{56}} & \textcolor{teal}{\ding{52}} & \textcolor{purple}{\ding{56}} & \textcolor{purple}{\ding{56}} & \textcolor{purple}{\ding{56}} & \textcolor{purple}{\ding{56}} & \textcolor{teal}{\ding{52}} & \textcolor{teal}{\ding{52}} & \textcolor{teal}{\ding{52}} & \textcolor{teal}{\ding{52}} \\
    \hspace{2mm}\ding{101} Trimmed Mean & \textcolor{purple}{\ding{56}} & \textcolor{purple}{\ding{56}} & \textcolor{purple}{\ding{56}} & \textcolor{purple}{\ding{56}} & \textcolor{purple}{\ding{56}} & \textcolor{teal}{\ding{52}} & \textcolor{purple}{\ding{56}} & \textcolor{purple}{\ding{56}} & \textcolor{purple}{\ding{56}} & \textcolor{purple}{\ding{56}} & \textcolor{purple}{\ding{56}} & \textcolor{purple}{\ding{56}} & \textcolor{purple}{\ding{56}} \\
    \hspace{2mm}\ding{101} Zero-knowledge proof & \textcolor{purple}{\ding{56}} & \textcolor{purple}{\ding{56}} & \textcolor{purple}{\ding{56}} & \textcolor{purple}{\ding{56}} & \textcolor{teal}{\ding{52}} & \textcolor{purple}{\ding{56}} & \textcolor{purple}{\ding{56}} & \textcolor{purple}{\ding{56}} & \textcolor{purple}{\ding{56}} & \textcolor{purple}{\ding{56}} & \textcolor{purple}{\ding{56}} & \textcolor{purple}{\ding{56}} & \textcolor{purple}{\ding{56}} \\
    \midrule
    \textbf{Attacks simulation} & \textcolor{purple}{\ding{56}} & \textcolor{purple}{\ding{56}} & \textcolor{teal}{\ding{52}} & \textcolor{purple}{\ding{56}} & \textcolor{orange}{\ding{117}} & \textcolor{purple}{\ding{56}} & \textcolor{purple}{\ding{56}} & \textcolor{purple}{\ding{56}} & \textcolor{purple}{\ding{56}} & \textcolor{purple}{\ding{56}} & \textcolor{purple}{\ding{56}} & \textcolor{orange}{\ding{117}} & \textcolor{purple}{\ding{56}} \\
    \midrule
    \hspace{2mm}\ding{101} Data Poisoning & \textcolor{purple}{\ding{56}} & \textcolor{purple}{\ding{56}} & \textcolor{teal}{\ding{52}} & \textcolor{purple}{\ding{56}} & \textcolor{teal}{\ding{52}} & \textcolor{purple}{\ding{56}} & \textcolor{purple}{\ding{56}} & \textcolor{purple}{\ding{56}} & \textcolor{purple}{\ding{56}} & \textcolor{purple}{\ding{56}} & \textcolor{purple}{\ding{56}} & \textcolor{teal}{\ding{52}} & \textcolor{purple}{\ding{56}} \\
    \hspace{2mm}\ding{101} Model poisoning & \textcolor{purple}{\ding{56}} & \textcolor{purple}{\ding{56}} & \textcolor{teal}{\ding{52}} & \textcolor{purple}{\ding{56}} & \textcolor{teal}{\ding{52}} & \textcolor{purple}{\ding{56}} & \textcolor{purple}{\ding{56}} & \textcolor{purple}{\ding{56}} & \textcolor{purple}{\ding{56}} & \textcolor{purple}{\ding{56}} & \textcolor{purple}{\ding{56}} & \textcolor{purple}{\ding{56}} & \textcolor{purple}{\ding{56}} \\
    \hspace{2mm}\ding{101} Byzantine & \textcolor{purple}{\ding{56}} & \textcolor{purple}{\ding{56}} & \textcolor{teal}{\ding{52}} & \textcolor{purple}{\ding{56}} & \textcolor{purple}{\ding{56}} & \textcolor{purple}{\ding{56}} & \textcolor{purple}{\ding{56}} & \textcolor{purple}{\ding{56}} & \textcolor{purple}{\ding{56}} & \textcolor{purple}{\ding{56}} & \textcolor{purple}{\ding{56}} & \textcolor{purple}{\ding{56}} & \textcolor{purple}{\ding{56}} \\
    \hspace{2mm}\ding{101} Label flipping & \textcolor{purple}{\ding{56}} & \textcolor{purple}{\ding{56}} & \textcolor{teal}{\ding{52}} & \textcolor{purple}{\ding{56}} & \textcolor{purple}{\ding{56}} & \textcolor{purple}{\ding{56}} & \textcolor{purple}{\ding{56}} & \textcolor{purple}{\ding{56}} & \textcolor{purple}{\ding{56}} & \textcolor{purple}{\ding{56}} & \textcolor{purple}{\ding{56}} & \textcolor{purple}{\ding{56}} & \textcolor{purple}{\ding{56}} \\
    \hspace{2mm}\ding{101} Backdoor & \textcolor{purple}{\ding{56}} & \textcolor{purple}{\ding{56}} & \textcolor{teal}{\ding{52}} & \textcolor{purple}{\ding{56}} & \textcolor{purple}{\ding{56}} & \textcolor{purple}{\ding{56}} & \textcolor{purple}{\ding{56}} & \textcolor{purple}{\ding{56}} & \textcolor{purple}{\ding{56}} & \textcolor{purple}{\ding{56}} & \textcolor{purple}{\ding{56}} & \textcolor{teal}{\ding{52}} & \textcolor{purple}{\ding{56}} \\
    \midrule
    \textbf{FL types} & \textcolor{orange}{\ding{117}} & \textcolor{teal}{\ding{52}} & \textcolor{teal}{\ding{52}} & \textcolor{orange}{\ding{117}} & \textcolor{teal}{\ding{52}} & \textcolor{teal}{\ding{52}} & \textcolor{orange}{\ding{117}} & \textcolor{teal}{\ding{52}} & \textcolor{orange}{\ding{117}} & \textcolor{teal}{\ding{52}} & \textcolor{teal}{\ding{52}} & \textcolor{orange}{\ding{117}} & \textcolor{teal}{\ding{52}} \\
    \midrule
    \hspace{2mm}\ding{101} Horizontal FL & \textcolor{teal}{\ding{52}} & \textcolor{teal}{\ding{52}} & \textcolor{teal}{\ding{52}} & \textcolor{teal}{\ding{52}} & \textcolor{teal}{\ding{52}} & \textcolor{teal}{\ding{52}} & \textcolor{teal}{\ding{52}} & \textcolor{teal}{\ding{52}} & \textcolor{teal}{\ding{52}} & \textcolor{teal}{\ding{52}} & \textcolor{teal}{\ding{52}} & \textcolor{teal}{\ding{52}} & \textcolor{teal}{\ding{52}} \\
    \hspace{2mm}\ding{101} Vertical & \textcolor{purple}{\ding{56}} & \textcolor{teal}{\ding{52}} & \textcolor{teal}{\ding{52}} & \textcolor{purple}{\ding{56}} & \textcolor{teal}{\ding{52}} & \textcolor{teal}{\ding{52}} & \textcolor{purple}{\ding{56}} & \textcolor{teal}{\ding{52}} & \textcolor{orange}{\ding{117}} & \textcolor{teal}{\ding{52}} & \textcolor{teal}{\ding{52}} & \textcolor{purple}{\ding{56}} & \textcolor{teal}{\ding{52}} \\
    \hspace{2mm}\ding{101} Split FL & \textcolor{purple}{\ding{56}} & \textcolor{purple}{\ding{56}} & \textcolor{teal}{\ding{52}} & \textcolor{purple}{\ding{56}} & \textcolor{purple}{\ding{56}} & \textcolor{purple}{\ding{56}} & \textcolor{purple}{\ding{56}} & \textcolor{teal}{\ding{52}} & \textcolor{purple}{\ding{56}} & \textcolor{purple}{\ding{56}} & \textcolor{orange}{\ding{117}} & \textcolor{purple}{\ding{56}} & \textcolor{purple}{\ding{56}} \\
    \midrule
    \textbf{Open source} & \textcolor{teal}{\ding{52}} & \textcolor{teal}{\ding{52}} & \textcolor{teal}{\ding{52}} & \textcolor{teal}{\ding{52}} & \textcolor{teal}{\ding{52}} & \textcolor{teal}{\ding{52}} & \textcolor{teal}{\ding{52}} & \textcolor{teal}{\ding{52}} & \textcolor{teal}{\ding{52}} & \textcolor{teal}{\ding{52}} & \textcolor{teal}{\ding{52}} & \textcolor{teal}{\ding{52}} & \textcolor{purple}{\ding{56}} \\
    \midrule
    \textbf{Documentation and tutorials} & \textcolor{teal}{\ding{52}} & \textcolor{teal}{\ding{52}} & \textcolor{teal}{\ding{52}} & \textcolor{teal}{\ding{52}} & \textcolor{teal}{\ding{52}} & \textcolor{teal}{\ding{52}} & \textcolor{orange}{\ding{117}} & \textcolor{teal}{\ding{52}} & \textcolor{orange}{\ding{117}} & \textcolor{orange}{\ding{117}} & \textcolor{teal}{\ding{52}} & \textcolor{teal}{\ding{52}} & \textcolor{orange}{\ding{117}} \\
    \bottomrule
  \end{tabular}}
\end{table*}

\textbf{CRYPTEN}~\cite{knott2021crypten}: \hl{CrypTen is a privacy-preserving ML framework implemented in Python and compatible with both Linux and Windows. Built on PyTorch, it provides MPC primitives, enabling collaborative computations on private data without exposing sensitive information. Its API closely resembles PyTorch, offering tensor computations, automatic differentiation, and modular neural networks, which simplify the integration of secure MPC techniques into ML workflows. CrypTen supports horizontal FL but lacks vertical and split FL capabilities. While it excels in secure aggregation and secret sharing, it does not implement DP or advanced attack simulations. The framework is open-source, well-documented, and user-friendly, making it accessible to ML practitioners. However, its reliance on an honest-but-curious threat model and limited support for advanced privacy mechanisms may restrict its application in certain adversarial scenarios.}

\textbf{FATE}~\cite{liu2021fate}: \hl{FATE is a flexible FL framework compatible with Linux and Windows, supporting popular programming languages like Python and Java. It offers comprehensive, secure computation protocols and diverse ML algorithms, including HE and MPC. FATE supports horizontal and vertical FL but lacks split FL capabilities. Its modular design provides end-to-end usability with pre-built components and user-friendly visualization tools, simplifying the implementation of privacy-preserving techniques. Additionally, FATE includes robust documentation, case studies, and tutorials to guide users. However, it does not implement DP or advanced attack simulations.} 

\textbf{FEDML}~\cite{he2020fedml, han2023fedmlsecurity, jin2023fedml}: \hl{FedML, also referred to as TensorOpera AI, is a versatile and robust FL platform compatible with Linux, macOS, and Windows, developed in Python. It supports three distinct computing paradigms: on-device training, distributed computing, and single-machine simulation, making it adaptable to various FL scenarios. FedML offers a flexible API design with comprehensive baseline implementations for optimizers, models, and datasets. Security and privacy are addressed through the FEDML-HE module, which employs HE techniques. Additionally, its FEDMLSecurity component includes FedMLAttacker for simulating all type of attacks and FedMLDefender for testing defensive strategies. It does not implement advanced privacy mechanisms such as Zero-knowledge proof or MPC. Despite these limitations, its extensive documentation and strong attack simulation features make it a recommended tool for research and practical applications in FL.}

\textbf{FEDSCALE}~\cite{lai2022fedscale}: \hl{FedScale is an FL benchmarking suite compatible with Linux, macOS, and Windows, implemented in Python. It provides scalable runtime and realistic datasets that support diverse FL tasks, such as image classification, object detection, and language modeling. Its high-level APIs simplify the implementation, deployment, and evaluation of FL algorithms, enabling researchers to benchmark FL at scale with minimal effort. FedScale employs DP techniques to enhance security and privacy but lacks support for other mechanisms. It supports horizontal FL but does not implement vertical or split FL. While its documentation is somewhat limited, it includes essential resources for experimentation.}

\textbf{FL AND DP}~\cite{rodriguez2020federated}: \hl{The Federated Learning (FL) and Differential Privacy (DP) framework is cross-platform, supporting Linux, Windows, and macOS, and is implemented in Python, Java, and C++. It emphasizes data privacy by integrating DP and holomorphic encryption techniques to quantify and mitigate privacy loss during distributed learning. The framework excels in ensuring privacy-preserving communication but lacks advanced security features. It supports horizontal and vertical FL but does not implement split FL. While the framework provides detailed documentation to guide users, its lack of a unified vision and a well-defined methodological workflow may limit its usability and effectiveness.}

\textbf{FLOWER}~\cite{beutel2020flower,li2021secure}: \hl{Flower is an open-source FL framework compatible with Linux, macOS, and Windows, implemented in Python. It is designed for large-scale FL experiments, supporting up to 15 million clients using only a pair of high-end GPUs, showcasing its scalability and efficiency. Flower excels in handling heterogeneous FL cross-device scenarios, making it suitable for diverse real-world applications. The framework prioritizes privacy by implementing various secure aggregation protocols, ensuring the server cannot inspect individual client models. However, it lacks support for advanced privacy techniques like FoolsGold or Geomed and does not include attack simulation features such as data poisoning or backdoor attacks. Flower supports horizontal and vertical FL but does not implement split FL. Detailed documentation and an active community enhance its usability by providing comprehensive guidance on installation, usage, and API references. Despite its limitations in attack simulations, Flower's scalability and flexibility make it a recommended tool for FL research and experimentation.}

\textbf{FLUTE}~\cite{hipolito2022flute}: \hl{The FLUTE (Federated Learning Under True Environment) framework is an open-source tool compatible with Linux, macOS, and Windows, implemented in Python (version 3.6 or higher). It is designed for high-performance FL research, enabling rapid prototyping and large-scale simulations of novel FL algorithms. FLUTE supports local and global DP methods, emphasizing data security and preservation. However, it lacks other advanced privacy mechanisms and does not include attack simulation features. While it supports horizontal FL, it does not implement vertical or split FL. The framework's documentation is available but incomplete, with no tutorials to assist new users.} 

\textbf{NVFLARE}~\cite{roth2022nvidia}: \hl{The NVIDIA FLARE (a.k.a NVFLARE) framework is a tool compatible with Linux and Windows. FLARE primarily supports Python as the programming language for developing FL workflows. Its exceptional features encompass state-of-the-art FL algorithms and approaches, allowing researchers to apply their data science workflows seamlessly using popular training libraries such as PyTorch, TensorFlow, XGBoost, or NumPy. Its lightweight, flexible, and scalable nature distinguishes the framework, rendering it suitable for real-world FL scenarios. It ensures secure and privacy-preserving multiparty collaboration by implementing HE or DP techniques. Comprehensive documentation provided by NVIDIA FLARE aids users in harnessing the framework's potential for both research and practical applications.}

\textbf{OPENFL}~\cite{foley2022openfl}: \hl{OpenFL is a Python-based FL framework compatible with Linux, macOS, and Windows. It supports developing and training ML and DL algorithms using TensorFlow, PyTorch, and other ML/DL frameworks, enhancing its adaptability for diverse use cases. As an open-source platform, OpenFL offers flexibility and customization options for researchers and developers. However, it lacks support for advanced privacy-preserving techniques such as secure aggregation or HE, limiting its security features. It supports horizontal FL but has an incomplete implementation of vertical FL. While documentation is available on its official website, it is incomplete and lacks specialized tutorials to guide new users effectively.}

\textbf{PaddleFL}~\cite{paddlefl}: \hl{PaddleFL is an open-source framework built on PaddlePaddle, supporting horizontal and vertical FL with privacy-preserving techniques like DP and Secure Aggregation. It is compatible with multiple platforms and languages but lacks split FL and attack simulation support. Despite its scalability, PaddleFL's usability is limited by sparse documentation and a predominantly Chinese-speaking community.}

\textbf{PYSYFT}~\cite{ziller2021pysyft}: \hl{PySyft is an FL library compatible with Linux, macOS, and Windows, primarily implemented in Python and extending popular DL frameworks like PyTorch. Its mission is to democratize privacy-preserving techniques in ML, making them accessible to researchers and data scientists. PySyft supports privacy-enhancing methods such as MPC and DP. However, it lacks advanced security protocols and does not provide attack simulation capabilities. PySyft implements horizontal vertical and an uncompleted version of split FL. Its comprehensive documentation includes detailed procedures, implementation guides, and example workflows, empowering users to effectively utilize the framework for privacy-focused FL projects.}

\textbf{TFF}~\cite{TFF, TFF_attacks}: \hl{TensorFlow Federated (TFF) is a Python-based framework that integrates with TensorFlow, supporting horizontal FL and incorporating privacy mechanisms like MPC and DP. It includes a simulation environment for testing attacks but lacks support for vertical and split FL and advanced privacy techniques like secure aggregation. Comprehensive documentation aids usability, though its limitations may restrict its use in highly adversarial settings.} 

\textbf{XFL}~\cite{wang2023xfl}: \hl{XFL is a versatile framework compatible with multiple platforms and languages, offering a user-friendly interface and pre-built algorithms for horizontal and vertical FL. It supports privacy-preserving techniques like HE, DP, and MPC but lacks support for split FL and attack simulations. While XFL simplifies deployment via Docker, incomplete documentation somewhat limits its usability.}

\emph{Lessons learned:} Given the previous details of each FL framework, FEDML emerges as the most complete solution since it incorporates many security and privacy methods, all the most common FL types, and vast documentation and tutorials. In addition, it highlights that since it is the only framework that includes a comprehensive suite of attack simulations. It is perfect for quickly testing new defenses proposed by security and privacy FL researchers. Nevertheless, FLOWER and FL AND DP are also relevant frameworks for researchers to consider due to the implementation of various security and privacy protocols. \textcolor{black}{However, they lack support for adversarial testing modules and fine-grained control over threat modeling, which limits their effectiveness for evaluating defenses under dynamic or adaptive adversaries. A notable gap is that none of the surveyed frameworks implement geometric median-based defenses like GeoMed, despite their empirical robustness against poisoning. In addition to missing implementations of advanced defenses (e.g., FLAME, Pruning-based defenses), we also find limited support for simulating realistic deployment conditions such as heterogeneous participation, client drift, or colluding Sybil attacks—factors increasingly relevant in real-world FL. We also observe that support for vertical and split FL is incomplete across most frameworks, which may hinder applications in finance, healthcare, or IoT, where data distributions are often partitioned by feature. Finally, the lack of standardized interfaces for measuring privacy-utility trade-offs (e.g., formal accounting of DP budgets vs. accuracy degradation) further limits reproducibility. Future framework development should prioritize modular threat modeling, integrated attack-defense testbeds, and support for adaptive, personalized, and hybrid defense strategies.}

\section{Main Applications}
\label{sec:applications}
We explore how various domains are employing this transformative FL technology. This chapter delves into real-world applications where FL plays a pivotal role, providing a deeper understanding of its practical significance. By highlighting relevant use cases, ranging from healthcare and finance to intrusion detection, we unveil the diverse scenarios where FL is making a substantial impact. \hl{Additionally, at the end of the section, we provide a paragraph of lessons learned based on analyzing the main applications of secure and private FL.}

\begin{figure}[t!]
  \centering
  \includegraphics[scale=0.35]{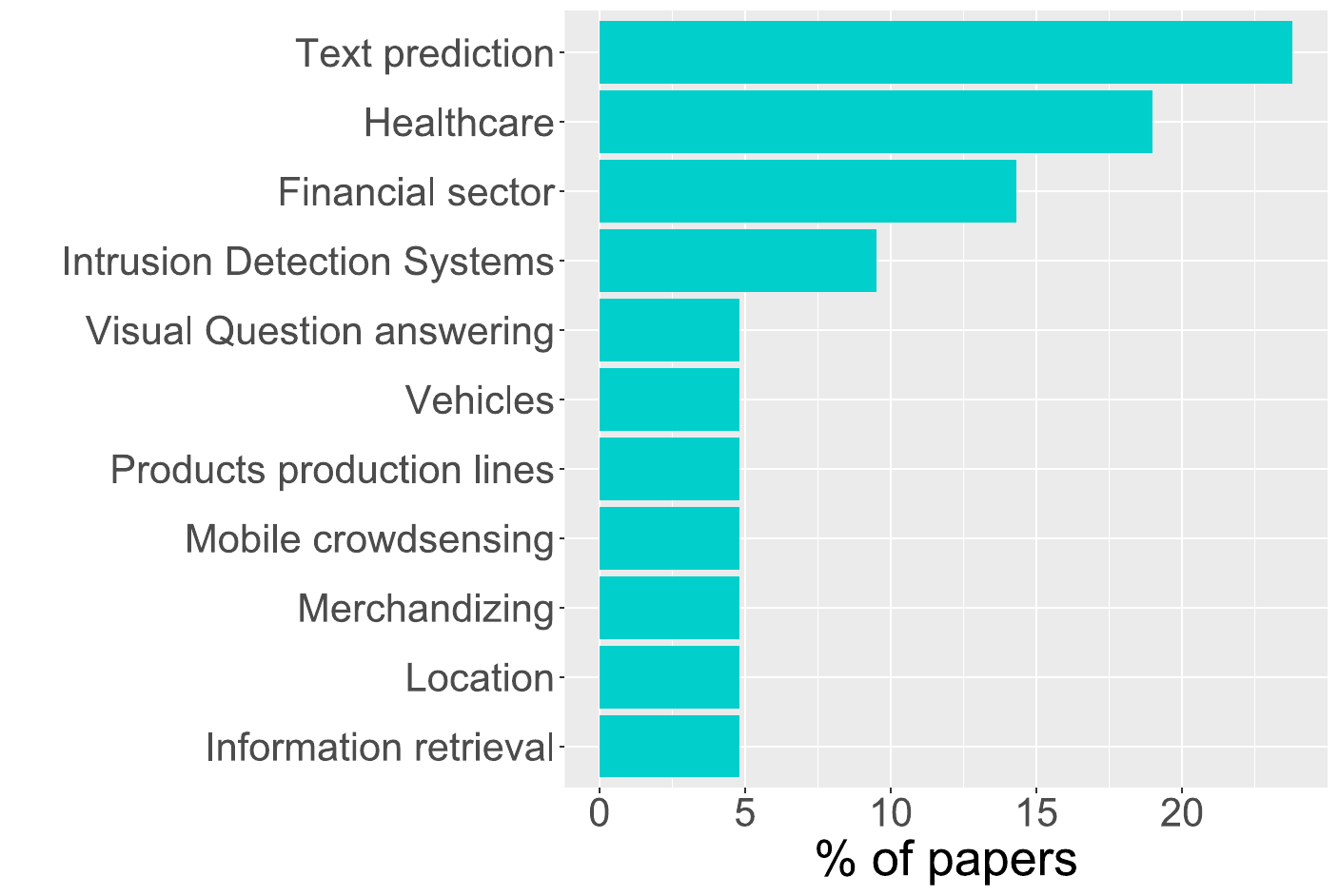}
  \caption{Main applications using privacy and secure FL}
  \label{fig:applications}
\end{figure}

Based on our literature review, we obtained the fields of applications that use FL in a private and secure context.Fig.~\ref{fig:applications} depicts the participation of each field over the papers analyzed. The top three fields applying secure and private FL are text prediction, healthcare, and the financial sector. The latter have been the most employed fields for a long time. Nevertheless, it is relevant to highlight that the intrusion detection systems field has gained strong participation among researchers in recent years. The following subsections describe how FL was employed for each field, emphasizing some challenges, attacks, and defenses utilized.

\paragraph{Text prediction}
Privacy and security are relevant in FL for text prediction because sensitive user data, such as personal messages or search queries, is processed locally on devices. Without strong privacy measures and secure communication (e.g., encryption), there is a high risk of exposing personal information, which could lead to breaches of user confidentiality or misuse of private data by malicious actors.
Advancements in FL for text prediction emphasize privacy and security through techniques like DP and local DP~\cite{batool2024secure,nagy2023privacy}. However, these methods often struggle to balance privacy and model performance, as stringent privacy measures can reduce prediction accuracy. Given the sensitive nature of textual data, ensuring security and privacy is vital for maintaining user trust and compliance with data protection regulations. Vulnerabilities in this field can stem from the decentralized nature of the data, with common attacks including poisoning attacks~\cite{birchman2024securing,nagy2023privacy}, GAN-based inference~\cite{wu2024backdoor,lycklama2023rofl}, and gradient inversion/suppression~\cite{liu2024raf,peng2024depth}. To address these risks, defenses such as DP, employed by Qi et al.~\cite{qi2023differentially}, can secure text data while preserving model utility.

\paragraph{Healthcare}
FL helps with privacy and security in healthcare by allowing institutions (i.e., hospitals) to train models collaboratively without sharing sensitive patient data. Techniques such as DP and HE safeguard patient information during model updates, addressing risks like data leakage and unauthorized access~\cite{mantey2024federated,lessage2024secure}. However, they can also introduce computational overhead and potentially compromise model accuracy. This domain encounters significant security and privacy challenges due to the sensitive nature of medical data. Common attacks include poisoning attacks~\cite{malisafe,sun2024fedkc,omran2023detecting}, where adversaries inject malicious data to undermine model integrity~\cite{zhang2023agrevader}, and gradient inversion/suppression~\cite{dao2024performance,li2022e2egi}, which attempts to recover private medical information from shared gradients~\cite{hatamizadeh2023gradient}. Membership inference attacks pose additional risks by revealing whether a specific patient was used in model training~\cite{sui2023multi}. 

To mitigate these threats, cryptographic techniques like HE and MPC ensure data privacy while allowing computations on encrypted data~\cite{muazu2024federated,singh2022framework}. DP also plays a crucial role in limiting the risk of sensitive information being memorized or inferred from model updates. Moreover, robust aggregation operators defend against poisoning and other adversarial manipulations, ensuring the integrity of the global model. By combining cryptographic techniques with privacy-preserving methods, healthcare FL systems can effectively protect against multifaceted attacks while maintaining accuracy and compliance with healthcare regulations, as highlighted by Singh et al.~\cite{hatamizadeh2023gradient}.

\paragraph{Financial sector}
FL enhances privacy and security in the financial sector by enabling institutions (i.e., banks and Fintech enterprises) to collaborate on fraud detection and risk management without sharing sensitive customer data. Ensuring security and privacy is critical in finance due to the sensitive nature of financial data and regulatory requirements, fostering customer trust and compliance with regulations like GDPR, which enable safer financial services. FL is particularly valuable for fraud detection and risk management but is susceptible to various attacks, notably GAN-based poisoning attacks~\cite{qiao2023privacy,wu2022federated}, which can degrade model performance and compromise privacy by manipulating training data, as highlighted by Qiao et al.~\cite{pang2022adi}. Moreover, MPC ensures that no single client gains access to sensitive financial data during joint model training.

\paragraph{Intrusion detection systems}
FL enhances privacy and security in intrusion detection systems by facilitating distributed model training without sharing raw logs. Techniques such as MPC and optimization-based input perturbation~\cite{chen2023feddef} guard against inference and poisoning attacks~\cite{yang2024adversarial}. However, challenges like deployment complexity and potential impacts on model accuracy persist. Security and privacy are vital in this domain, as adequate intrusion detection safeguards sensitive environments from unauthorized access, ensuring compliance with security standards and fostering user trust in system reliability. In IoT networks, FL encounters significant security challenges, including label-flipping attacks~\cite{lavaur2024systematic}, where malicious clients manipulate labels to mislead the global model, as Yang et al.\cite{chen2023feddef} noted.

\paragraph{Visual question-answering}
In visual question-answering (VQA) models using FL, several security and privacy concerns arise due to the complexity of the task and the diverse data types involved. One prominent attack in vertical FL VQA is the ADI, where adversaries manipulate input data, such as images, to dominate the learning process and reduce the contributions of other clients, as explored by Pang et al.~\cite{pang2022adi}. GAN-based inference is another risk, where attackers attempt to reconstruct private information, such as images or questions, from model updates. Anomaly detection can be employed to defend against these threats by identifying and excluding manipulated data or adversarial inputs before they influence the model. DP can obscure sensitive images or question details to protect clients' local data~\cite{li2020oscar}. Additionally, robust aggregation operators ensure that adversarial contributions, like poisoned data, do not degrade the overall model performance. Vertical FL VQA systems can leverage these defenses to maintain privacy and security, enabling collaborative model training without exposing sensitive visual or textual information.

\paragraph{Vehicles}
Under this field, using FL with secure and private defenses is critical because connected cars generate sensitive data about drivers' locations, routes, driving behaviors, and vehicle diagnostics, and protecting this information prevents unauthorized tracking, behavior profiling, and potential safety vulnerabilities. In the vehicle field, ADI attacks, such as random or bounded mutation, can manipulate vehicle data and degrade model performance, as Pang et al.~\cite{pang2022adi} reported. Additionally, model inconsistency may arise from adversarial updates across clients. To defend against these, robust aggregation operators reduce the impact of malicious updates, while DP and additive noise protect sensitive vehicle data from being inferred through model updates. These defenses ensure secure and accurate FL models in vehicle-related tasks.

\paragraph{Products production line}

In this field, privacy and security in FL are functional because manufacturers can collaboratively improve their production models and optimize processes while securely keeping sensitive proprietary data (like manufacturing parameters, quality control metrics, and production recipes) within their facilities, preventing industrial espionage and maintaining competitive advantages. FL faces label-flipping and backdoor attacks in product production lines, which can compromise model accuracy and reliability in assembly processes~\cite{li2021byzantine}. To counter these threats, robust aggregation operators filter out harmful contributions from adversaries, while anomaly detection identifies and excludes suspicious data. Additionally, DP protects sensitive production metrics during model training~\cite{mangal2016using}. 

\paragraph{Mobile crowd-sensing}
Privacy and security in mobile crowd-sensing FL are crucial to protect sensitive location and behavioral data, preventing unauthorized tracking and identity breaches while enabling valuable insights for urban planning and services. FL is susceptible to eavesdropping and membership inference attacks in mobile crowd-sensing, which can compromise client privacy. Additionally, poisoning attacks can manipulate model updates, degrading performance~\cite{zhao2022crowdfl}. To defend against these threats, secure aggregation methods based on the threshold Paillier cryptosystem protect model confidentiality while DP obscures individual contributions. Robust aggregation operators also help mitigate the impact of adversarial updates.

\paragraph{Merchandising}
Privacy and security in FL regarding this field are relevant because retailers handle sensitive customer purchasing patterns, inventory strategies, and pricing data across multiple locations. Thus, protecting such information prevents competitors from accessing valuable business intelligence while fostering peer-to-peer modeling to optimize merchandising decisions and customer experience across store networks.
In this field, FL faces threats like poisoning attacks, where malicious clients corrupt the global model by manipulating their local data, and membership inference attacks, which attempt to deduce the presence of specific data samples in the training dataset~\cite{zhang2023agrevader}. To defend against these attacks, robust aggregation operators can filter out adversarial updates, ensuring that only reliable contributions influence the global model. Additionally, employing DP techniques helps obscure individual shopping histories, protecting sensitive merchandising data from exposure~\cite{shokri2017membership}. 

\paragraph{Location}
FL's privacy and security are paramount in location services as they protect users' sensitive movement patterns and visited places while leveraging multiparty learning to improve location-based services without exposing individual data. FL is vulnerable to location tracking attacks in the location field, where adversaries attempt to infer users' movements or patterns from shared model updates. Membership inference attacks can also expose sensitive information about individuals based on their location check-ins~\cite{zhang2023agrevader}. Implementing DP techniques can obscure individual check-in data to mitigate these threats, protecting user privacy while allowing practical model training. Additionally, robust aggregation operators can help filter out adversarial contributions, ensuring that only trustworthy data influences the global model~\cite{shokri2017membership}. 

\paragraph{Information retrieval}
In information retrieval, FL's privacy measures protect users' sensitive search patterns and interests while permitting collaborative improvement of search systems without exposing personal data. In this field, FL faces challenges such as insufficient training data, where individual users may lack enough interactions to achieve high search effectiveness. The latter can be exploited through model inversion attacks, where adversaries infer sensitive user data from shared model parameters~\cite{wang2023analysis}. DP techniques can be employed to protect individual search interactions, ensuring user privacy while still allowing model training. Additionally, robust aggregation operators can help mitigate the effects of malicious updates, enhancing the reliability of the global model.

\emph{Lessons learned:} The analysis of secure and private FL across main domains highlights its transformative potential and persistent challenges. The widespread adoption in fields like text prediction, healthcare, and finance underscores the necessity of FL for protecting sensitive user data while enabling collaborative model training. However, ensuring privacy and model performance remains a central challenge, as strict privacy mechanisms often introduce accuracy and computational efficiency trade-offs. The rise of FL in intrusion detection systems and mobile crowd-sensing indicates an increasing awareness of its role in security-sensitive environments, though these applications face threats like poisoning attacks and adversarial data manipulation. Across all domains, the decentralized nature of FL introduces vulnerabilities such as gradient inversion and membership inference attacks, emphasizing the need for robust defense mechanisms. Notably, emerging vehicles, manufacturing, and information retrieval applications demonstrate FL's adaptability yet reveal unique domain-specific risks, from adversarial attacks in autonomous driving to industrial espionage in production lines. \textcolor{black}{Another key limitation is the lack of standardized, reproducible evaluation pipelines across domains, making it difficult to compare defense effectiveness or understand trade-offs across threat models. Furthermore, many application areas lack publicly available benchmarks that reflect realistic attack scenarios, hindering the development and validation of domain-adaptive security mechanisms.} A key lesson is that while FL enhances data privacy, both its privacy and security largely depend on continuous advancements in cryptographic techniques, adversarial defenses, and efficient aggregation strategies tailored to each field's requirements.

\section{Future Directions}
\label{sec:future_work}
While significant strides have been made in addressing FL's security and privacy challenges, several areas remain ripe for exploration and improvement. The complexity and evolving nature of FL environments necessitate ongoing research to refine existing techniques and develop novel solutions. This section outlines vital areas for future work, highlighting the need for advanced methods to enhance the robustness of FL systems against emerging threats. It emphasizes the importance of addressing limitations in current approaches and exploring innovative strategies that balance security, privacy, and efficiency. 

\subsection{Security Future Directions}

Security in FL remains a significant challenge, especially in light of sophisticated poisoning and backdoor attacks. Future directions should focus on developing robust and adaptive security mechanisms that can detect and mitigate these threats while maintaining the integrity of the global model. The emphasis would be on improving the resilience of FL systems, enhancing verification processes, and developing scalable solutions that support high performance even in adversarial settings.

\begin{itemize} \item \textbf{Enhanced Robustness Against Advanced Attacks:} As discussed in Section~\ref{sec:security}, FL environments face significant challenges from adversarial attacks, such as model poisoning and backdoor insertion, particularly in heterogeneous data settings~\cite{yang2024robust}.

\hl{To counter model poisoning attacks, where compromised clients degrade global model performance, future work should explore adaptive aggregation techniques that dynamically adjust the contributions of client updates based on anomaly detection metrics. For example, methods like adaptive local aggregation}~\cite{zhang2023fedala}\hl{ or sparsification-based defenses}~\cite{panda2022sparsefed}\hl{ could be extended to incorporate real-time monitoring of update trajectories}~\cite{ma2021asynchronous}\hl{. Additionally, integrating client-side defenses like FL-WBC, which perturbs parameter spaces affected by attacks, could mitigate long-term attack impacts}~\cite{sun2021fl}.

\hl{For backdoor attacks, in which malicious clients insert triggers into models to induce targeted misclassifications, future defenses could leverage hybrid anomaly detection approaches combining statistical gradient analysis and cryptographic verification}~\cite{mi2023identifying}\hl{. Techniques like ARIBA have shown promise in identifying distributional anomalies in model updates. Furthermore, incorporating multi-method adaptive aggregation algorithms (e.g., SAPAA-MMF) could enhance robustness by balancing contributions based on data quality and variance}~\cite{zhang2024federated}\hl{. Exploring interdisciplinary methods inspired by biological immune systems could also provide novel insights for adaptive and self-healing mechanisms in FL systems.}

\item \textbf{Resilience to Emerging Threats in Dynamic Environments:}
Deploying FL in dynamic settings such as autonomous vehicles and smart cities introduces unique vulnerabilities, including free-riding attacks, model extraction attacks, and jamming threats~\cite{ma2024stability}. Addressing these challenges requires targeted strategies:

\emph{Free-Riding Attacks:} Free-riders exploit FL aggregation protocols by contributing no meaningful updates while benefiting from the global model~\cite{fraboni2021free}. Future work should explore advanced anomaly detection mechanisms such as high-dimensional clustering techniques (e.g., STD-DAGMM) to identify free-riders~\cite{zhu2021advanced}. Integrating blockchain-based accountability frameworks could also enhance trust by recording client contributions transparently~\cite{asif2023integrating}.

\hl{\emph{Model Extraction Attacks:} Malicious clients can reverse-engineer global models to steal intellectual property or compromise privacy}~\cite{khan2024fed}\hl{. Hybrid encryption techniques combining HE and MPC could be employed to counter this. Furthermore, gradient obfuscation methods that distort shared parameters without degrading model performance warrant investigation}~\cite{yue2023gradient}.

emph{Collaborative Jamming Attacks:} Jamming attacks in FL-based 5G networks disrupt communication channels, degrading model performance. In 5G networks, jammers exploit public NR standards (e.g., PUCCH intra-slot hopping patterns and RACH protocols)~\cite{arjoune2020smart} to disrupt synchronization signals with energy-efficient methods like reactive jamming, posing critical risks to public safety and military operations. Mitigation includes spread spectrum techniques (DSSS/FHSS) and ML-based detection (XGBoost ensembles achieving 99.72\% accuracy). Concurrently, model extraction attacks—enabled via API query duplication (e.g., LLM "leeching")~\cite{hou2025privacy}—threaten proprietary models in finance and healthcare. Defenses like ModelGuard’s information-theoretic perturbation maintain <3\% utility loss while thwarting extraction. A promising direction involves implementing decentralized jamming detection frameworks using convolutional autoencoders for unsupervised anomaly detection and FedProx algorithms for supervised classification~\cite{kuili2025two}.

\item \textbf{Fairness, Bias Mitigation, and Security Integration:} As FL models are increasingly deployed in sensitive applications like healthcare and finance, ensuring fairness under adversarial conditions remains a critical challenge. \hl{A particularly concerning threat is fairness attacks (poisoning), where attackers manipulate data or model updates to introduce or amplify bias}~\cite{wang2023pass}\hl{. These attacks disproportionately harm specific groups, such as racial minorities or underrepresented communities, making fairness a direct target of adversarial manipulation.

Future research should focus on developing fairness-preserving aggregation methods that integrate anomaly detection with fairness constraints. For example, leveraging techniques like FairFed}~\cite{ezzeldin2023fairfed}\hl{, which adaptively reweights client contributions based on fairness metrics, could counteract biased updates. Additionally, interdisciplinary approaches combining cryptographic tools with fairness-aware algorithms can enhance defenses against malicious clients}~\cite{du2021fairness}\hl{. Addressing indirect bias--where even non-malicious clients contribute biased data unintentionally--requires advanced techniques like fairness-aware pruning or incorporating domain adaptation methods to balance performance across heterogeneous client distributions}~\cite{yang2023csra}.

\item \textbf{Scalable, Efficient, and Verifiable Secure Aggregation:} As FL scales to larger and more complex systems, secure aggregation techniques face significant challenges, particularly in mitigating attacks such as model poisoning and Sybil attacks. \hl{A critical technical challenge is designing aggregation protocols that balance computational efficiency with robust security guarantees. For instance, lightweight encryption mechanisms like homomorphic hash functions or single-mask symmetric encryption could reduce overhead while maintaining privacy and verifiability}~\cite{zhang2021faithful}\hl{. Additionally, dynamic masking strategies, which adapt to threat levels in real-time, could enhance resilience against predictable attack patterns}~\cite{xiong2024privmaskfl}.

\hl{To counter Sybil's attacks effectively, interdisciplinary approaches integrating DP with anomaly detection methods show promise. For example, combining DP with graph-based anomaly detection could identify malicious clients based on their interaction patterns}~\cite{kong2024federated}\hl{. Furthermore, verifiable aggregation protocols such as LightVeriFL can ensure the integrity of updates by leveraging homomorphic commitment schemes for lightweight verification}~\cite{buyukates2024lightverifl}. Future work should explore these approaches in scenarios with high user dropout rates to ensure robustness.

Blockchain technology offers a promising avenue for tamper-resistant and auditable aggregation. However, its scalability remains a concern due to high computational costs. A potential solution is hybrid architectures that combine blockchain with adversarial training techniques or verifiable delay functions to balance security and efficiency~\cite{cai2024esvfl}. Another plausible solution is using dedicated off-chain servers to handle validation and aggregation (e.g., Fantastyc's proof generation), reducing on-chain operations by 70\%~\cite{boitier2024fantastyc}. Moreover, reinforcement learning-based adaptive Proof-of-Work (PoW) dynamically adjusts mining difficulty in response to real-time miner capabilities and network conditions, reducing energy waste by up to 45\% and lowering computational overhead for honest clients~\cite{sethi2023reinforcement}. Research should focus on optimizing these solutions for decentralized FL settings where central servers are absent.

\end{itemize}

\subsection{Privacy Future Directions}

Privacy preservation is a critical aspect of FL, mainly when dealing with sensitive data distributed across multiple clients. The future of privacy-enhancing techniques will focus on improving the efficiency and scalability of existing methods, making them suitable for a wide range of applications, from edge devices to large-scale cross-silo FL environments. The goal is to ensure data privacy without compromising model performance or significantly increasing computational and communication overhead.

\begin{itemize}
    \item \textbf{Privacy-Enhancing Techniques for Non-IID Data:} Non-IID data presents one of the biggest challenges in FL, particularly in safeguarding privacy while ensuring robust model performance (see Section~\ref{sec:non-iid-impact}). \hl{In non-IID scenarios, privacy-preserving techniques like HE, MPC, and DP face limitations due to data heterogeneity, which increases susceptibility to targeted inference attacks}~\cite{he2024privacy}\hl{. Attackers can exploit discrepancies in data distributions across clients to perform data reconstruction or membership inference attacks. To address these vulnerabilities, future research should focus on:

    \emph{Dynamic Privacy Mechanisms:} Developing adaptive DP mechanisms that adjust privacy budgets based on client-specific data heterogeneity. For instance, privacy budgets could be dynamically allocated using metrics such as the Hellinger, Jensen-Shannon, or Earth mover's distances to quantify inter-client distribution disparities}~\cite{jimenez2024fedartml}\hl{.
    
    \emph{Scalable Encryption Protocols:} Optimizing HE and MPC for non-IID settings by reducing computational overhead through techniques like hybrid encryption schemes or gradient compression}~\cite{elkordy2022basil}\hl{.
    
    \emph{Robust Aggregation Methods:} Designing aggregation techniques that mitigate the influence of skewed client updates, such as similarity-weighted aggregation or clustering-based approaches}~\cite{he2024privacy}.

    \item \textbf{Integration of Advanced Cryptographic Protocols:} As FL continues to scale, particularly in sensitive domains like IoT and healthcare, privacy remains vulnerable to specific attacks such as inference and canary gradient attacks. \hl{Future research must focus on integrating advanced cryptographic protocols that enhance privacy while minimizing performance costs. Such future work includes the following technical challenges: 
    
    \emph{Inference Attacks:} Adversaries reconstruct sensitive data from aggregated model updates. This requires efficient HE schemes that support secure aggregation without significant computational overhead}~\cite{han2023fedmlsecurity}\hl{.
    
    \emph{Canary Gradient Attacks:} Attackers inject small perturbations into gradients or weight updates. Existing cryptographic methods struggle to detect such subtle manipulations}~\cite{cao2022mpaf}\hl{.
    
    \emph{Key Management in HE:} Current single-key HE schemes risk key leaks, necessitating multi-key or secret-sharing schemes for enhanced security}~\cite{wang2025privacy}\hl{.

    Therefore, some proposed solutions that can be explored in future research are: First, develop hybrid cryptographic frameworks combining HE with MPC to protect against both classical and quantum adversaries}~\cite{han2023fedmlsecurity}\hl{.
    Second, implement adaptive gradient clipping techniques alongside DP to mitigate inference and canary attacks without degrading model accuracy}~\cite{munoz2023survey}\hl{.
    Third, design decentralized key management systems using secret sharing or blockchain-based approaches to enhance security in HE implementations.}
    
    \item \textbf{Enhanced Verification of Aggregated Models:} Ensuring the integrity of aggregated models while preserving privacy is critical, especially given the growing threat of GAN-based inference attacks. \hl{These attacks exploit GANs to infer sensitive information about training data in FL settings}~\cite{morettin2024unified}\hl{. Future research must address specific challenges, such as reducing computational overhead and communication costs while maintaining robust privacy guarantees. Promising directions include:

    \emph{Lightweight Verifiable Aggregation Protocols:} Techniques such as homomorphic hashing and bilinear aggregate signatures have shown potential for verifying aggregation results}~\cite{guo2020verifl}\hl{. However, these methods often face scalability issues due to high-dimensional model gradients. Research should optimize these protocols by leveraging advanced cryptographic techniques like polynomial commitments or ProxyZKP frameworks}~\cite{li2024polynomial}\hl{.
    
    \emph{Combating GAN-Based Attacks:} Defense mechanisms like Anti-GAN frameworks, which manipulate visual features to thwart GAN-based inference attacks, are promising}~\cite{xing2023zero}\hl{. Future work could explore integrating such frameworks with secure aggregation techniques to enhance privacy without compromising model accuracy.

    \emph{ZKPs for Privacy-Preserving Verification:} ZKPs enable entities to prove the correctness of computations without revealing sensitive data}~\cite{xing2023zero}\hl{. While ZKPs hold great promise for FL, current implementations face scalability challenges. 

    \emph{Optimizing ZKP Scalability:} Techniques like zk-SNARKs and zk-STARKs provide efficient proof systems but require further optimization for large-scale FL applications. The ProxyZKP framework, which uses polynomial decomposition to reduce proof generation times, offers a viable path forward}~\cite{li2024polynomial}. Another avenue is using Batch verification processes to verify multiple proofs simultaneously, cutting verification overhead by up to 70\%, while recursive composition hierarchically aggregates proofs into compact representations, ideal for large-scale deployments~\cite{he2015accountable}. For resource-constrained environments, collaborative zk-SNARKs distribute proof generation across parties, linearly reducing per-node complexity~\cite{liu2024scalable}.

    \hl{\item \textbf{Quantitative Privacy-Performance Trade-off Models:} While numerous studies have explored the qualitative trade-offs between privacy, security, and model performance in FL, a unified quantitative framework remains an open challenge. Existing research provides valuable insights into individual trade-offs, such as the impact of privacy budgets in DP on model utility, but lacks a standardized mathematical formulation that systematically captures these interdependencies. Future research should focus on developing mathematical models that integrate privacy loss, computational overhead, and model accuracy into a single framework. These models could incorporate utility functions that balance security guarantees with performance metrics, similar to approaches in economic game theory or optimization-based frameworks}~\cite{mohammadi2024balancing}\hl{. By addressing these directions, future research can bridge the gap between qualitative discussions and rigorous quantitative analysis, ensuring a more precise understanding of privacy-performance trade-offs in FL.}

\end{itemize}

\subsection{Joint Directions on Security–Privacy in FL}

\textcolor{black}{While this survey separately addresses challenges in security and privacy, real-world FL deployments often suffer from their combined vulnerabilities. A critical future direction lies in understanding how attacks on one axis may amplify risks on the other, and how certain defenses may inadvertently open new threat vectors. For instance, a poisoning attack can manipulate the model's sensitivity to benign client gradients, thereby increasing the effectiveness of gradient inversion techniques. Similarly, colluding Sybil clients can bias the global model toward a specific user’s data distribution, enhancing the attacker’s chances in subsequent membership inference. Conversely, privacy defenses like secure aggregation or heavy DP noise may hinder the detection of adversarial behavior, thereby weakening overall system security.}

\textcolor{black}{Future research should pursue co-designed mechanisms that bridge this gap:}
\begin{itemize}
    \item \textcolor{black}{Integrated threat modeling that considers both privacy leakage and security compromise in unified scenarios.}
    
    \item \textcolor{black}{Privacy-aware robust aggregation techniques that maintain anomaly detection capabilities even under DP noise.}
    
    \item \textcolor{black}{Layer-wise defense strategies that protect sensitive layers with cryptographic tools while preserving transparency in others for anomaly auditing.}
    
    \item \textcolor{black}{Client-side collaborative monitoring, using lightweight trusted execution environments (TEEs) to audit gradients locally before encrypted aggregation.}
    
    \item \textcolor{black}{Benchmark frameworks that evaluate FL systems against compound attack scenarios rather than isolated vectors.}
\end{itemize}

\textcolor{black}{Based on the previous analysis, FL research must evolve from siloed views to holistic frameworks, ensuring that strengthening one defense front does not unintentionally weaken the other. Tackling this interplay remains a fundamental challenge and opportunity for building resilient and trustworthy federated systems.}

\textcolor{black}{An emerging dimension of this security–privacy interplay arises from the integration of FL with Generative AI (GenAI) systems, including large language models (LLMs). While these models offer powerful personalization and collaborative capabilities, they also amplify both axes of vulnerability. For example, GenAI systems are particularly prone to data memorization, making them susceptible to privacy leakage even under secure aggregation \cite{yan2024protecting}. This problem becomes more pronounced in FL, where attackers may exploit intermediate gradients or personalized prompts to extract or reconstruct client data. This raises new privacy risks beyond what traditional FL defenses like secure aggregation or DP were designed to handle. Security threats also take new forms. For instance, poisoning attacks in generative models may bias completions toward specific ideologies or inject imperceptible toxic content. Detection and mitigation become more complex when the goal of an attack is to subtly influence output distributions rather than flip classification labels \cite{alber2025medical}. Additionally, verifying the integrity and alignment of decentralized GenAI systems becomes increasingly difficult without centralized auditing. Future research must thus explore new FL frameworks tailored for generative tasks. These may include federated instruction tuning pipelines with private prompt alignment, hybrid split-federated architectures to manage compute imbalance, and adaptive privacy controls that account for generative memorization risks, for example, considering \textit{federated unlearning} notions \cite{liu2024survey}. Addressing these questions is essential for deploying GenAI responsibly in distributed environments, as well as for advancing the robustness and trustworthiness of FL systems more broadly.}

\section{Conclusion}
\label{sec:conclusion}
This survey provides an in-depth analysis of the security and privacy challenges in FL. It reveals that despite FL's design to enhance data privacy, it is susceptible to various threats, such as data poisoning, model inversion, and backdoor attacks, underscoring the need for effective defense mechanisms. By categorizing these attacks and their impacts, we offer a structured understanding of FL systems' diverse threats. We also highlight the importance of balancing privacy, security, and model performance through techniques like cryptographic methods and DP. Recent research trends indicate a growing focus on addressing these issues, calling for scalable and adaptive solutions suitable for dynamic environments. Future research should develop innovative, energy-efficient solutions to address the identified challenges, paving the way for more secure and practical FL applications. Overall, this survey is a valuable resource for future work advancing secure, privacy-preserving collaborative learning systems.

\section{Acknowledgments}
\noindent Daniel Mauricio Jimenez G. was partially supported by
PNRR351 TECHNOPOLE -- NEXT GEN EU Roma Technopole -- Digital Transition,
FP2 -- Energy transition and digital transition in urban regeneration
and construction and Sapienza Ateneo Research grant ``La
disintermediazione della Pubblica Amministrazione: il ruolo della
tecnologia blockchain e le sue implicazioni nei processi e nei ruoli
della PA.''.
José Luis Hernández Ramos was supported by a Leonardo
Grant 2023 to Researchers and Cultural Creators from the BBVA Foundation, a Consolidación Investigadora grant (CNS2023-145059) funded by MICIU/AEI/10.13039/501100011033 and by the European Union NextGenerationEU/PRTR, the project ONOFRE-4 funded by MICIU/AEI/10.13039/501100011033 and FEDER, UE, and the project PCI2023-145989-2 (REMINDER)  funded by MICIU/AEI/10.13039/501100011033 and the European Union. 
Aris Anagnostopoulos was supported by
the ERC Advanced Grant 788893 AMDROMA, the EC H2020RIA project
``SoBigData++'' (871042), the PNRR MUR project PE0000013-FAIR, the PNRR
MUR project IR0000013-SoBigData.it, and the MUR PRIN project 2022EKNE5K
``Learning in Markets and Society.''.
Ioannis Chatzigiannakis and Yelizaveta Falkouskaya were supported by
PE07-SERICS (Security and Rights in the Cyberspace) -- European Union
Next-Generation-EU-PE0000014 (Piano Nazionale di Ripresa e Resilienza -- PNRR).
Andrea Vitaletti was supported by  PE11 - MICS (Made in Italy --
Circular and Sustainable) -- European Union Next-Generation-EU
(Piano Nazionale di Ripresa e Resilienza -- PNRR).

\bibliographystyle{elsarticle-num-names}
\bibliography{13-references}

\end{document}